\documentclass[12pt, centerh1]{article}

\usepackage{graphicx,colonequals}
\usepackage{url}
\usepackage{amsmath}
\usepackage{amsfonts}
\usepackage{amssymb,amsthm}
\usepackage{url}
\usepackage{color}
\usepackage{booktabs,natbib}
\usepackage{amssymb,multirow,booktabs,graphicx}
\usepackage{graphics,psfrag,amsmath,natbib,amsthm,xr}
\usepackage{dsfont,latexsym,epsfig,colonequals}
\usepackage[titletoc,toc,title]{appendix}
\usepackage{color}

\newcommand{\bx}{\boldsymbol{x}}

\newcommand{\bz}{\boldsymbol{z}}

\newcommand{\bm}{(\boldsymbol{x}_i - \boldsymbol{\mu})}

\newcommand{\bv}{\boldsymbol{v}}

\newcommand{\bw}{\boldsymbol{w}}
\newcommand{\bA}{\boldsymbol{A}}
\newcommand{\bD}{\boldsymbol{D}}

\newcommand{\bJ}{\boldsymbol{J}}

\newcommand{\bM}{\boldsymbol{M}}
\newcommand{\bX}{\boldsymbol{X}}

\newcommand{\bW}{\boldsymbol{W}}
\newcommand{\bB}{\boldsymbol{B}}

\newcommand{\bI}{\boldsymbol{I}}
\newcommand{\bQ}{\boldsymbol{Q}}

\newcommand{\bV}{\boldsymbol{V}}
\newcommand{\bZ}{\boldsymbol{Z}}

\newcommand{\cL}{\mathcal{L}}

\newcommand{\EE}{\mathbb{E}}

\newcommand{\bmu}{\mbox{\boldmath $\mu$}}
\newcommand{\blambda}{\mbox{\boldmath $\lambda$}}

\newcommand{\bbeta}{\mbox{\boldmath $\beta$}}

\newcommand{\bSigma}{\mbox{\boldmath $\Sigma$}}
\newcommand{\bGamma}{\mbox{\boldmath $\Gamma$}}
\newcommand{\bLambda}{\mbox{\boldmath $\Lambda$}}

\newcommand{\bTheta}{\mbox{\boldmath $\Theta$}}
\newcommand{\bDelta}{\mbox{\boldmath $\Delta$}}

\newcommand{\new}{\mbox{\tiny new}}

\newcommand{\tr}{\operatorname{tr}}
\newcommand{\veco}{\operatorname{vec}}

\begin{document}

\title{Mixtures of Multivariate Power Exponential Distributions}
\author{Utkarsh J.\ Dang\thanks{Department of Biology, McMaster University, Hamilton, Ontario L8S-4L8, Canada. E-mail: udang@mcmaster.ca}, Ryan P.\ Browne\thanks{Department of Mathematics \& Statistics, McMaster University, Hamilton, Ontario L8S-4L8, Canada.}, and Paul D.\ McNicholas\thanks{Department of Mathematics \& Statistics, McMaster University, Hamilton, Ontario L8S-4L8, Canada.}}
\date{ }
\maketitle

\begin{abstract}
An expanded family of mixtures of multivariate power exponential distributions is introduced. While fitting heavy-tails and skewness has received much attention in the model-based clustering literature recently, we investigate the use of a distribution that can deal with both varying tail-weight and peakedness of data. A family of parsimonious models is proposed using an eigen-decomposition of the scale matrix. A generalized expectation-maximization algorithm is presented that combines convex optimization via a minorization-maximization approach and optimization based on accelerated line search algorithms on the Stiefel manifold. Lastly, the utility of this family of models is illustrated using both toy and benchmark data.
\end{abstract}

\section{Introduction}
\label{sec:introduction}

Mixture models have become the most popular methodology to investigate heterogeneity in data \citep[cf.][]{titterington1985, mclachlan2000}. Model-based learning makes use of mixture models to partition data points. Model-based clustering and classification refer to the scenarios where observations have no known labels and some known labels, respectively, \textit{a~priori}.  The number of these partitions or clusters may or may not be known in advance. 
While approaches based on mixtures of Gaussian distributions \citep[e.g.,][]{banfield1993, celeux1995} remain popular for model-based clustering, these algorithms are susceptible to performing poorly in the presence of outliers. As a result, more robust mixtures of distributions are becoming increasingly popular. Some of these mixtures aim to tackle tail-weight \citep[e.g.,][]{Andr:McNi:Exte:2011,andrews2012, forbes2013}, some deal with skewness \citep[e.g.,][]{lin2007b, franczak2014}, while others account for both \citep[e.g.,][]{Karl:Sant:Mode:2009,subedi2014, vrbik2014, browne2013}. 

Herein, we utilize a family of mixture models based on the multivariate power exponential (MPE) distribution \citep{gomez1998}. This distribution is sometimes also called the multivariate generalized Gaussian distribution. Depending on the shape parameter $\beta$, two kinds of distributions can be obtained: for $0<\beta<1$ a leptokurtic distribution is obtained, which is characterized by a thinner peak and heavy tails compared to the Gaussian distribution; whereas, for $\beta>1$, a platykurtic distribution is obtained, which is characterized by a flatter peak and thin tails compared to the Gaussian distribution. The distribution is quite flexible: for $\beta=0.5$, we have a Laplace (double-exponential) distribution and, for $\beta=1$, we have a Gaussian distribution. Furthermore, when $\beta \rightarrow \infty$ the MPE becomes a multivariate uniform distribution.
\begin{figure*}
\centering
\includegraphics[trim = 0mm 30mm 0mm 155mm, clip, scale=0.9,angle=270]{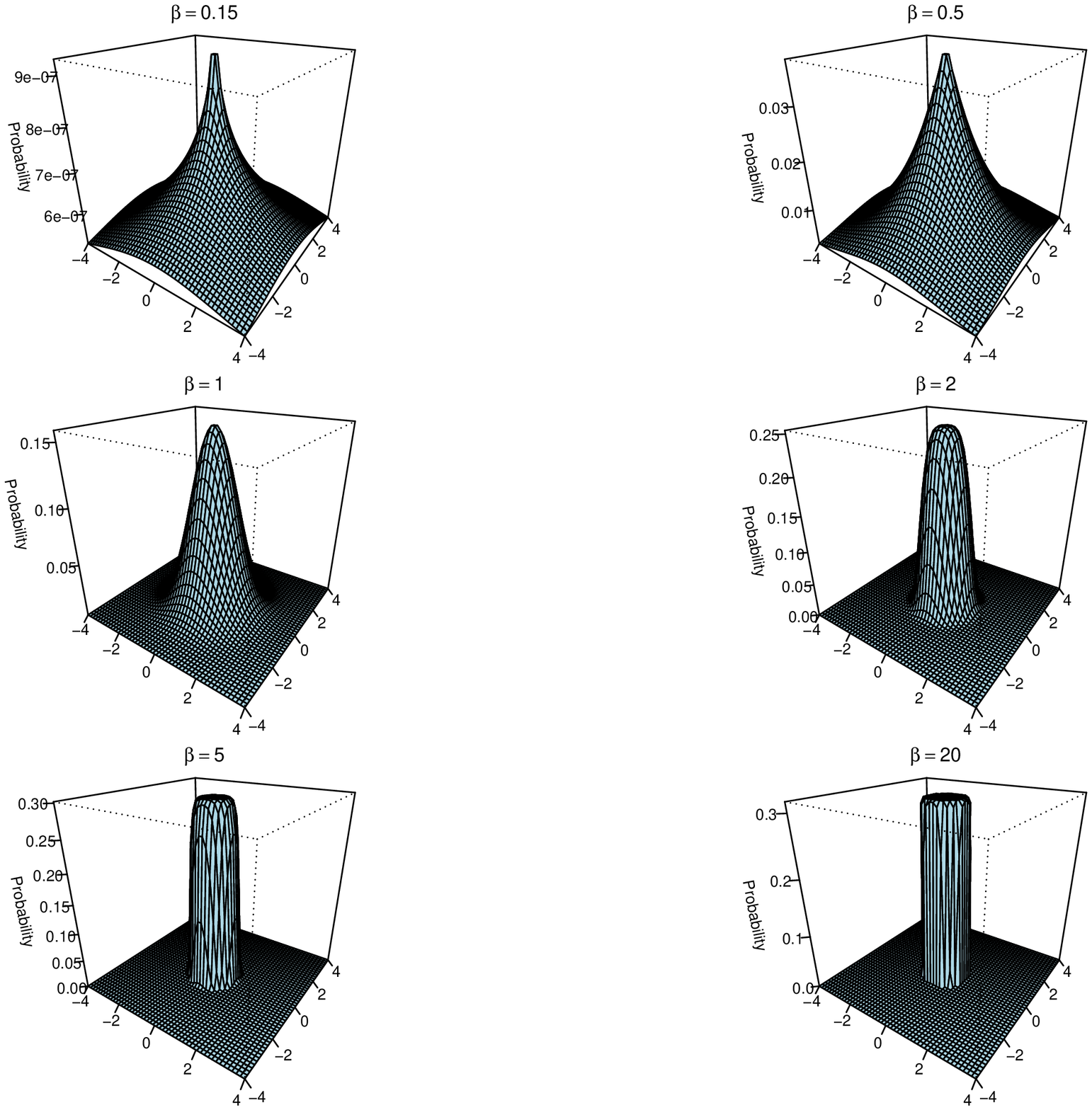}
\hspace{-0.6in}
\includegraphics[trim = 0mm 155mm 0mm 30mm, clip, scale=0.9,angle=270]{plotswithdiffbetas.eps}
\caption{Density plots for different values of $\beta$. The MPE distribution is quite flexible: for $\beta=0.5$, we have a Laplace (double-exponential) distribution and for $\beta=1$, we have a Gaussian distribution. Furthermore, as $\beta \rightarrow \infty$, the MPE distribution becomes a multivariate uniform distribution.}
\label{diffbetas}
\end{figure*}

The MPE distribution has been used in many different applications \citep{lindsey1999, cho2005, verdoolaege2008}. However, due to difficulties in estimating the covariance over the entire support of the shape parameter $\beta\in(0,\infty)$, its potential has not yet been fully explored. This distribution presents a difficult parameter estimation problem because none of the parameter estimates are available in closed form. 
Previously proposed estimation strategies have included optimization based on geodesic convexity for unconstrained covariance matrices \citep{zhang2013} and Newton-Raphson recursions \citep{pascal2013}.
Some work with this distribution has focused on the special case where $0<\beta<1$ \citep{gomez2008, bombrun2012, pascal2013}. However, for imposing parsimony in a traditional model-based clustering context (through different constraints on terms of specific decompositions of the component scale, or covariance, matrices), these methods are not ideal. Previously, a family of five models based on mixtures of MPE distributions has been used for robust clustering \citep{zhang2010}. This work made use of fixed point iterations for the special case where $0<\beta<2$ (see Appendix \ref{sec:Zhangcomparison}). Within $0 <\beta < 2$, the fixed point algorithm converges; however, it yields monotonic improvements in log-likelihood only for $0 <\beta \leq 1$. For $\beta \geq 2$, this fixed point algorithm is guaranteed to diverge, which leads to (negative) infinite log-likelihood values. 

Herein, a generalized expectation-maximization \citep[GEM;][]{dempster1977} strategy is proposed and illustrated. This algorithm works for $0<\beta<\infty$. This estimation procedure also guarantees monotonicity of the log-likelihood. We make use of MM algorithms \citep{hunter2000} and accelerated line search algorithms on the Stiefel manifold \citep{absil2009,browne2014}. This allows for the estimation of a wide range of constrained models, and a family of sixteen MPE mixture models is presented. These models can account for varying tail weight and peakedness of mixture components. In Section \ref{sec:mpe}, we summarize the MPE distribution. Section \ref{sec:mpeinference} gives a GEM algorithm for parameter estimation. Section \ref{sec:mperesults} investigates the performance of the family of mixture models on toy and benchmark data. We conclude with a discussion and suggest some avenues for further research in Section \ref{sec:mpediscussion}.

\section{Multivariate Power Exponential Distribution} \label{sec:mpe}

A random vector $\bX$ follows a $p$-dimensional power exponential distribution \citep{landsman2003} if the density is of the form
\begin{equation} \label{mpegeneral}
h(\bx|\bmu,\bSigma,r,s)=c_p|\bSigma|^{-\frac{1}{2}} \exp\left\{ -\frac{r}{2^s}\delta(\bx)^s \right\},
\end{equation}
where $$c_p=\frac{s\Gamma\left(\frac{p}{2}\right)}{(2\pi)^{p/2}\Gamma\left(\frac{p}{2s}\right)} r^{p/(2s)},$$ $\delta(\bx):=\delta \left(\bx|\bmu,\bSigma \right)=\left (\bx-\bmu \right)' \bSigma^{-1} \left(\bx-\bmu \right)$, $\bmu$ and $\bSigma$ are the location parameter (also the mean) and positive-definite scale matrix, respectively, and $r, s>0$. This elliptical distribution is a multivariate Kotz-type distribution. However, it has identifiability issues concerning $\bSigma$ and $r$: the density with $\bTheta=\{\bmu,\bSigma^*,r^*,s\}$, where $\bSigma^*=\bSigma/2$ and $r^*=r/2^s$, is the same as \eqref{mpegeneral}.

Using the parametrization given by \cite{gomez1998}, a random vector $\bX$ follows a $p$-dimensional power exponential distribution if the density is
\begin{equation} \label{mpedef}
f(\bx|\bmu,\bSigma,\beta)=k|\bSigma|^{-\frac{1}{2}} \exp\left\{ -\frac{1}{2} \delta(\bx)^\beta \right\},
\end{equation}
where $$k=\frac{p\Gamma\left(\frac{p}{2}\right)}{\pi^{p/2}\Gamma\left(1+\frac{p}{2\beta}\right)2^{1+\frac{p}{2\beta}}},$$ $\delta(\bx):=\delta \left(\bx|\bmu,\bSigma \right)=\left (\bx-\bmu \right)' \bSigma^{-1} \left(\bx-\bmu \right)$, $\bmu$ is the location parameter (also the mean), $\bSigma$ is a positive-definite scale matrix, and $\beta$ determines the kurtosis. 
Moreover, it is a special parameterization of the MPE distribution given in \eqref{mpegeneral}, with $r=2^{\beta-1}$ and $s=\beta$. The covariance and multidimensional kurtosis coefficient for this distribution are 
\begin{equation} \label{covariance}
\text{Cov}(\bX)=\frac{2^{1/\beta}\Gamma\left(\frac{p+2}{2\beta}\right)}{p\Gamma \left(\frac{p}{2\beta} \right)}\bSigma
\end{equation}
and
\begin{equation} \label{kurtosis}
\gamma_2(\bX)=\frac{p^2\Gamma\left(\frac{p}{2\beta}\right) \Gamma\left(\frac{p+4}{2\beta}\right)}{\Gamma^2\left(\frac{p+2}{2\beta}\right)}-p(p+2),
\end{equation}
respectively \citep{gomez1998}. Here, $\gamma_2(\bX)$ denotes the multidimensional kurtosis coefficient that is defined as $$\EE\left\{ \left[(\bX-\bmu)'\text{Var}(\bX)^{-1}(\bX-\bmu)\right]^2\right\}-p(p+2)$$ \citep{mardia1980,gomez1998}. For $\beta \in (0,1)$, the MPE distribution is a scale mixture of Gaussian distributions \citep{gomez2008}. 

Based on the MPE distribution, a mixture model can conveniently be defined as $$g(\bx|\bTheta)=\sum_{g=1}^G \pi_g f\left(\bx| \bmu_{g}, \bSigma_{g},\beta_g\right),$$ 
where $f(\cdot)$ is the $g$th component density and $\bTheta$ denotes all parameters. Here, $\bmu_g$, $\bSigma_g$, and $\beta_g$ denote the mean, scale matrix, and shape parameter, respectively, of the $g$th component. Here, $\pi_1,\ldots,\pi_G$ are the mixing weights such that $\pi_{g}>0$ ($g = 1,\ldots,G$) and $\sum^G_{g=1}\pi_{g}=1$. Note that mixtures of MPE distributions have previously been shown to be identifiable \citep{zhang2010}.

Because the number of parameters in the scale matrix increases quadratically with data dimensionality, it is common practice to impose a decomposition that allows for reduction in the number of parameters to be estimated. An eigen-decomposition decomposes a component covariance matrix into the form $\bSigma_g=\lambda_g {\bGamma_g} {\bDelta_g} {\bGamma_g}'$, where $\lambda_g$, $\bGamma_g$, and $\bDelta_g$ can be interpreted geometrically \citep{banfield1993}. Specifically, $\bDelta_g$ is a diagonal matrix with entries proportional to the eigenvalues of $\bSigma_g$ (with $|\boldsymbol{\Delta}_g|=1$), $\lambda_g$ is the associated constant of proportionality, and $\boldsymbol{\Gamma}_g$ is a $p\times p$ orthogonal matrix of the eigenvectors of~$\bSigma_{g}$ (with entries ordered according to the eigenvalues). Constraining these terms to be equal or variable across groups allows for a family of fourteen parsimonious mixture models \citep{celeux1995}. In this paper, we work with a subset of eight parsimonious models (EII, VII, EEI, VVI, EEE, EEV, VVE, and VVV), including the most parsimonious (EII) and the fully unconstrained (VVV) models (Table \ref{tab:models}). In addition, there is the option to constrain $\beta_g$ to be equal across groups. This option, together with the covariances structures, results in a family of sixteen models. The nomenclature for this family is a natural extension of that used for the covariance structures, e.g., the model with a VVI scale structure and $\beta_g$ constrained to be equal across groups is denoted VVIE. This family of models is referred to as the ePEM (eigen-decomposed power exponential mixture) family hereafter.

\begin{table*}
\caption{Nomenclature, scale matrix structure, and the number of free scale parameters for the ePEM family of models.} \label{tab:models}
\begin{tabular*}{1.0\textwidth}{@{\extracolsep{\fill}}lllllr}
\toprule
Model & $\lambda_g$ & $\bDelta_g$ & $\bGamma_g$ & $\boldsymbol{\Sigma}_g$ & Free Cov.\ Parameters \\
\midrule
EII & Equal    & Spherical & -            & $\lambda \boldsymbol{I}$           & 1\\
VII & Variable & Spherical & -            & $\lambda_g \boldsymbol{I}$         & $G$\\[2mm]
EEI & Equal    & Equal     & Axis-Aligned & $\lambda \boldsymbol{\Delta}$      & $p$\\
VVI & Variable & Variable  & Axis-Aligned & $\lambda_g \boldsymbol{\Delta}_g$  & $Gp$\\[2mm]
EEE & Equal    & Equal     & Equal        & $\lambda\boldsymbol{\Gamma}\boldsymbol{\Delta}\boldsymbol{\Gamma}'$  & $p\left(p+1\right)/2$\\
EEV & Equal    & Equal     & Variable     & $\lambda\boldsymbol{\Gamma}_g\boldsymbol{\Delta}\boldsymbol{\Gamma}_g'$  & $Gp(p+1)/2 - (G-1)p$ \\
VVE & Variable & Variable  & Equal        & $\lambda_g\boldsymbol{\Gamma}\boldsymbol{\Delta}_g\boldsymbol{\Gamma}'$  & $p(p+1)/2 + (G-1)p$ \\
VVV & Variable & Variable  & Variable   & $\lambda_g \boldsymbol{\Gamma}_g \boldsymbol{\Delta}_g \boldsymbol{\Gamma}_g'$  & $Gp\left(p+1\right)/2$ \\
\bottomrule
\end{tabular*}
\bigskip
\end{table*}

\section{Inference}\label{sec:mpeinference}
The expectation-maximization (EM) algorithm \citep{dempster1977} is an iterative procedure based on the complete-data likelihood. At each iteration, the the expected value of the complete-data log-likelihood is maximized to yield updates for the parameters of interest. The expectation-conditional-maximization (ECM) algorithm \citep{meng1993} replaces the maximization step of the EM algorithm with a number of conditional maximization (CM) steps. This might be necessary due to the form of the likelihood or because the conditional maximization steps are less computationally expensive. In our parameter estimation algorithm, CM steps are used within a framework that increases, rather than maximizes, the expected value of the complete data log-likelihood at each iteration. Such an approach, i.e., one that has the latter feature, is called a GEM algorithm.
The parameter updates associated with our GEM algorithm are given in Appendix \ref{sec:parupdates}. 

\section{Results}\label{sec:mperesults}
For our numerical analyses, we use the Bayesian information criterion \citep[BIC;][]{schwarz1978} and the integrated complete likelihood \citep[ICL;][]{biernacki2000} for model selection. A stopping criterion based on the Aitken acceleration \citep{aitken1926} is used to determine convergence and the adjusted Rand index \citep[ARI;][]{hubert1985} 
is used for performance assessment. More details are in Appendix \ref{sec:modelspecifications}. In Appendix~\ref{sec:Zhangcomparison}, we compare the performance of our algorithm to an algorithm based on fixed point iterations.

\subsection{Simulations}\label{sec:simulations}
For simulating from the MPE distribution, a modified version of the function {\tt rmvpowerexp} from package {\tt MNM} \citep{MNM} in {\sf R} \citep{R2013} is used. The function was modified due to a typo in the {\tt rmvpowerexp} code. This program utilizes the stochastic representation of the MPE distribution \citep{gomez1998} to generate data. This works quite well in lower dimensions. In higher dimensions, a Metropolis-Hastings-based simulation rule can easily be constructed. We illustrate 
the performance of our family of models using simulations in a wide range of scenarios: for light-tailed components, for light- and heavy-tailed components, for data simulated from Gaussian and $t$-distributions, for higher-dimensional data, and for low overall sample size. When data are simulated from the MPE distribution only, we also show parameter recovery. For comparison to existing mixture models based on elliptically contoured distributions, the {\tt mixture} \citep{mixture} and {\tt teigen} \citep{teigen} packages in {\sf R} are employed. These packages implement mixtures of Gaussian and mixtures of multivariate Student-t distributions, respectively. To facilitate a direct comparison, we restrict {\tt mixture} and {\tt teigen} to the analogues of the ePEM models (Table~\ref{tab:models}). Note that we use the {\tt mixture} package rather than {\tt mclust} \citep{fraley2012} because the VVE model is available within {\tt mixture} but not within {\tt mclust}, which only implements ten of the 14 models of \cite{celeux1995}. Moreover, as compared to {\tt Rmixmod} \citep{rmixmod}, certain models in the {\tt mixture} family are better optimized for higher dimensions \citep[cf.][]{browne2014mm}. Note that the {\tt teigen} package additionally allows for constraining of the degrees of freedom parameter ($\nu$). Hence, a VVIV model implies that $\lambda_g$, $\bDelta_g$, and $\nu_g$ are different between groups, and $\bGamma_g$ is the identity matrix. Note that the same starting values are used for all three algorithms, i.e., for each $G$, the initial $\tau_{ig}$ are selected from the best $k$-means clustering results from ten random starting values for the $k$-means algorithm \citep{hartigan1979}. 

\paragraph{Simulation 1: Two light-tailed components}
A two-component mixture is simulated with 450 observations with the sample sizes for each group sampled from a binomial distribution with success probability $0.45$. The first component is simulated from a two-dimensional MPE distribution with zero mean, identity scale matrix, and $\beta_1=2$. The second component is simulated from a two-dimensional MPE distribution with mean $(2,0)'$, identity scale matrix, and $\beta_2=5$. Note that this corresponds to an EIIV model. The simulated components are not well separated. All three algorithms are run on 100 such data sets. For the ePEM family, a two-component model is selected by the BIC (and the ICL) for each of the 100 data sets. On the other hand, for the {\tt mixture} family, the BIC selects a two-component model 77 times, and three, four, and five component models are selected 15, 6, and 2 times, respectively. Similarly, for the {\tt teigen} family, two, three, four, and five component models are selected 61, 10, 26, and 3 times, respectively. Clearly, for both of the latter families, more components are being fitted to deal with the light-tailed nature of the data.

\begin{figure*}[!h]
\centering
\includegraphics[scale=0.55,angle=270]{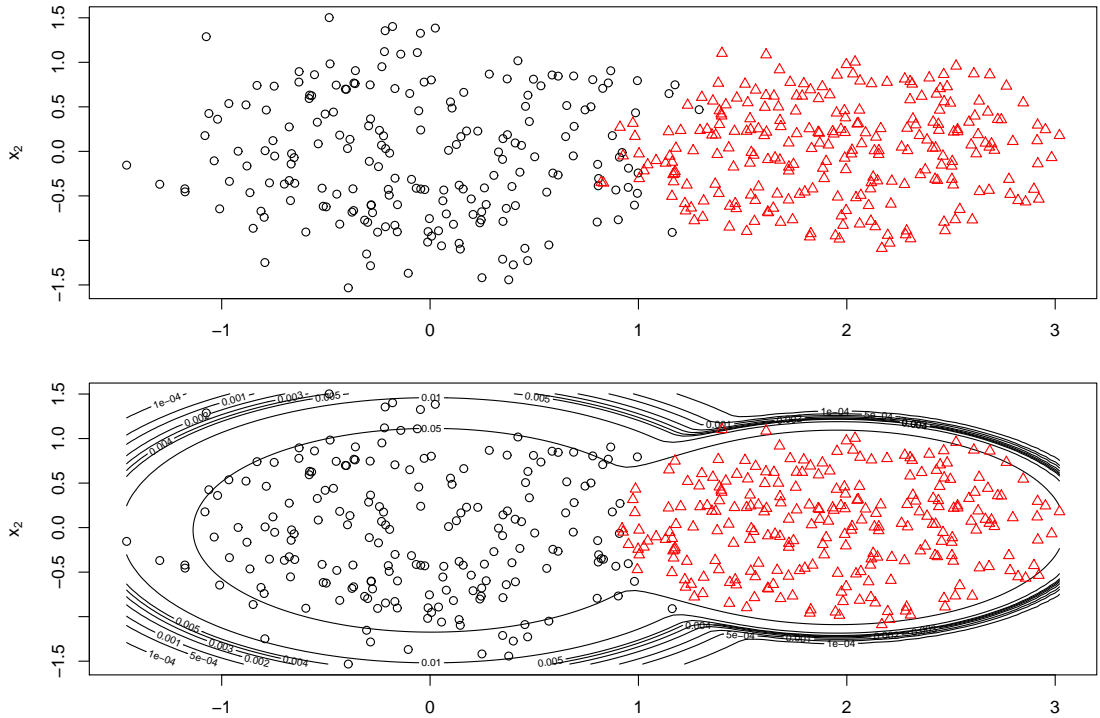}
\caption{Plots showing the generated data (top) and the fitted density (bottom) using the selected model from the ePEM family for Simulation 1. This figure appears in color in the electronic version of this article.}
\label{sim1contour}
\end{figure*}

For the ePEM family, the EIIV model is selected by the BIC 97 times out of 100, with the VIIE model selected the other 3 times. The ARI values for the selected ePEM models range from 0.81 to 0.95, with a median (mean) ARI value of 0.88 (0.88). The selected {\tt mixture} models yield ARI values ranging between 0.30 and 0.96, with a median (mean) value of 0.85 (0.79). Similarly, the {\tt teigen} family yields ARI values ranging between 0.29 and 0.94, with a median (mean) value of 0.80 (0.69). A contour plot shows the fit of a selected EIIV model to an example data set (Figure~\ref{sim1contour}). The estimated mean, variance (using \eqref{covariance}), and $\bbeta$ are given in Table \ref{mpeparestsim1}. Clearly, the estimates are quite close to the true parameter values.
\begin{table*}[!ht]
\caption{True parameter values along with mean and standard deviations of the parameter estimates (rounded off to 2 decimals) for the selected model from the 100 runs for Simulation 1.} \label{mpeparestsim1}
\smallskip
\begin{tabular*}{1.0\textwidth}{@{\extracolsep{\fill}}lrrr}
\hline
Parameter & True values & Mean estimates & Standard deviations\\
  \hline
$\pi_1$ & 0.45 & 0.45 & 0.03\\
$\pi_2$ & 0.55 & 0.55 & 0.03\\ 
$\bmu_1$ & $(0, 0)'$ & $(-0.01, 0.00)'$ & $(0.05, 0.04)'$\\
$\bmu_2$ & $(2, 0)'$ & $(2.00, -0.00)'$ & $(0.03, 0.02)'$\\
$\text{Var}_1$ & 0.40 & 0.40 & 0.02 \\
$\text{Var}_2$ & 0.28 & 0.28 & 0.01 \\
$\beta_1$ & 2 & 2.10 & 0.39\\
$\beta_2$ & 5 & 5.77 & 3.06\\
\hline
\end{tabular*}
\bigskip
\end{table*}

The impact of multiple initializations in terms of the model and number of components selected is also evaluated. Here, the $k$-means initialization mentioned above is repeated 25 times for all 100 simulated data sets. In all cases, the same model is selected (by the BIC) for all 25 runs. Hence, hereafter, only one $k$-means initialization (as explained in Section~\ref{sec:simulations}) is used for all simulated and real data.

\paragraph{Simulation 2: Light and heavy-tailed components}
A three-component mixture is simulated with 500 observations in total. Group sample sizes are sampled from a multinomial distribution with mixing proportions $(0.35, 0.15, 0.5)'$. The first component is simulated from a 3-dimensional MPE distribution with mean $(0, 2, 0)'$ and $\beta_1=0.85$. The second component is simulated from a 3-dimensional MPE distribution with mean $(2, 5, 0)'$ and $\beta_2=3$. Lastly, the third component is simulated from a 3-dimensional MPE distribution with mean $(4, 2, 0)'$ and $\beta_3=5$. To generate the scale matrices (using an EEEV scale structure), we use $$\bGamma_1=\bGamma_2=\bGamma_3=\begin{pmatrix}
  0.36&0.48&-0.8 \\	
  -0.8&0.6&0\\
  0.48&0.64&0.6
 \end{pmatrix},$$ $\bDelta_1=\bDelta_2=\bDelta_3=\text{diag}(4, 3, 1)$, where $\text{diag}(\cdot)$ refers to a diagonal matrix. 
 
\begin{figure}[!h]
\centering
\includegraphics[trim=10mm 10mm 0mm -10mm,clip,scale=0.6,angle=270]{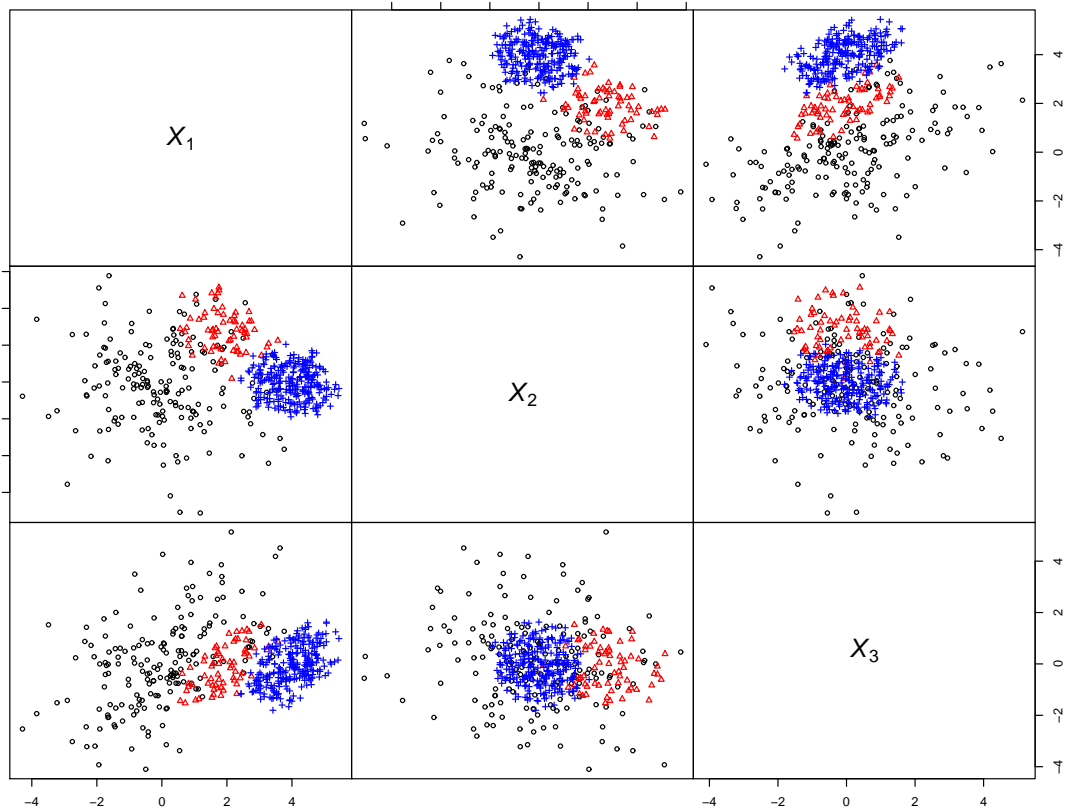}
\caption{Scatter plots showing an example of a generated data set for Simulation~2.}
\label{sim2}
\end{figure}

For all three families, the BIC selects a three-component model for each of the 100 runs. For the ePEM family, the BIC selects an EEEV (VVEE) model 99 (1) times. The ARI values for the selected models from the {\tt mixture} family range between 0.87 and 0.96 with a median (mean) value of 0.92 (0.92). Similarly, the {\tt teigen} family yields ARI values between 0.85 and 0.96 with a median (mean) value of 0.92 (0.91). Even though all three families select the same number of components every time, the estimated ARI values for the selected ePEM models are higher, ranging between 0.91 and 0.98 with a median (mean) value of 0.94 (0.94). A scatter plot showing an example of the generated data is given in Figure \ref{sim2}. The estimated mean, covariance, and $\bbeta$ are given in Table \ref{mpeparestsim2}. 

\begin{table*}[!ht]
\caption{True parameter values along with mean and standard deviations of the parameter estimates (rounded off to 2 decimals) for the selected model from the 100 runs for Simulation 2.} \label{mpeparestsim2}
\smallskip
\begin{tabular*}{1.0\textwidth}{@{\extracolsep{\fill}}lrrr}
\hline
Parameter & True values & Mean estimates & Standard deviations\\
  \hline
$\pi_1$ & 0.35 & 0.35 & 0.02\\
$\pi_2$ & 0.15 & 0.15 & 0.02\\ 
$\pi_3$ & 0.5 & 0.5 & 0.02\\ 
$\bmu_1$ & $(0, 2, 0)'$ & $(-0.02, 1.97, 0.01)'$ & $(0.12, 0.18, 0.14)'$\\
$\bmu_2$ & $(2, 5, 0)'$ & $(1.99, 4.98, 0.00)'$ & $(0.08, 0.12, 0.10)'$\\
$\bmu_3$ & $(4, 2, 0)'$ & $(4.00, 2.00, 0.01)'$ & $(0.03, 0.04, 0.03)'$\\
$\text{Covariance}_1$ & $\begin{pmatrix}
2.86 & -0.45 & 1.75 \\ 
-0.45 & 5.64 & -0.59 \\ 
1.75 & -0.59 & 3.89 \\ 
  \end{pmatrix}$ & $\begin{pmatrix}
2.88 & -0.41 & 1.75 \\ 
-0.41 & 5.71 & -0.57 \\ 
1.75 & -0.57 & 3.89 \\   
\end{pmatrix}$ & $\begin{pmatrix}
0.27 & 0.16 & 0.20 \\ 
0.16 & 0.53 & 0.18 \\ 
0.20 & 0.18 & 0.34 \\ 
  \end{pmatrix}$ \\
  $\text{Covariance}_2$ & $\begin{pmatrix}
0.49 & -0.08 & 0.30 \\ 
-0.08 & 0.97 & -0.10 \\ 
0.30 & -0.10 & 0.67 \\  
  \end{pmatrix}$ & $\begin{pmatrix}
0.49 & -0.07 & 0.30 \\ 
-0.07 & 0.98 & -0.10 \\ 
0.30 & -0.10 & 0.67 \\ 
\end{pmatrix}$ & $\begin{pmatrix}
0.05 & 0.03 & 0.04 \\ 
0.03 & 0.09 & 0.03 \\ 
0.04 & 0.03 & 0.06 \\ 
  \end{pmatrix}$ \\
  $\text{Covariance}_3$ & $\begin{pmatrix}
0.42 & -0.07 & 0.26 \\ 
-0.07 & 0.83 & -0.09 \\ 
0.26 & -0.09 & 0.57 \\ 
  \end{pmatrix}$ & $\begin{pmatrix}
0.42 & -0.06 & 0.25 \\ 
-0.06 & 0.83 & -0.08 \\ 
0.25 & -0.08 & 0.57 \\ 
\end{pmatrix}$ & $\begin{pmatrix}
0.02 & 0.02 & 0.02 \\ 
0.02 & 0.04 & 0.02 \\ 
0.02 & 0.02 & 0.03 \\ 
  \end{pmatrix}$ \\
$\beta_1$ & 0.85 & 0.87 & 0.17\\
$\beta_2$ & 3 & 3.49 & 1.22\\
$\beta_3$ & 5 & 5.93 & 1.37\\
\hline
\end{tabular*}
\bigskip
\end{table*}

\paragraph{Simulation 3: Higher-dimensional data} Here, parameter recovery is illustrated for the ePEM family on higher dimensional data. One-hundred samples of a thirty dimensional two-component mixture model are simulated in the fashion of \cite{murray2014}. Group sample sizes are sampled from a binomial distribution with success probability $0.35$ and an overall sample size of 400. The first component is simulated from a 30-dimensional MPE distribution with zero mean. The second component is simulated from a 30-dimensional MPE distribution with mean $(3, 3, 3)' \otimes \boldsymbol{1}_{10}$, where $\boldsymbol{1}_{10}$ denotes a column vector of length 10 with all entries equalling 1. The common scale matrix is generated using $$\begin{pmatrix}
  1&0.1&0.2 \\	
  0.1&1.5&0.3\\
  0.2&0.3&1.2
 \end{pmatrix} \otimes \boldsymbol{I}_{10},$$
where $\boldsymbol{I}_{10}$ denotes a 10-dimensional identity diagonal matrix. The recovered parameter estimates are found to be close on average to the generating parameters. Due to the dimensionality, we follow \cite{murray2014} and report the Frobenius norms of the biases of the parameter estimates in Table \ref{mpeparestsim3}. Clearly, the estimated parameters are quite close to the generating parameters. Note that while the purpose of this simulation is to investigate parameter estimation in higher dimensions, all 100 runs yield perfect clustering.

\begin{table*}[!ht]
\caption{True parameter values along with the Frobenius norms of the biases of the parameter estimates (rounded off to 2 decimals) for the selected model from the 100 runs for Simulation 3.} \label{mpeparestsim3}
\smallskip
\begin{tabular*}{1.0\textwidth}{@{\extracolsep{\fill}}lrr}
\hline
Parameter & True values & $\|\text{Bias}\|$\\
  \hline
$\pi_1$ & 0.35 & 0.00\\
$\pi_2$ & 0.65 & 0.00\\ 
$\bmu_1$ & $(0, 0, 0)' \otimes \boldsymbol{1}_{10}$ & 0.02\\
$\bmu_2$ & $(3, 3, 3)' \otimes \boldsymbol{1}_{10}$ & 0.05\\
$\text{Covariance}_1$ & $\frac{2^{1/\beta_1}\Gamma\left(\frac{p+2}{2\beta_1}\right)}{p\Gamma \left(\frac{p}{2\beta_1} \right)} \times\begin{pmatrix}
  1&0.1&0.2 \\	
  0.1&1.5&0.3\\
  0.2&0.3&1.2
 \end{pmatrix} \otimes \boldsymbol{I}_{10}$ & 0.26\\
 $\text{Covariance}_2$ & $\frac{2^{1/\beta_2}\Gamma\left(\frac{p+2}{2\beta_2}\right)}{p\Gamma \left(\frac{p}{2\beta_2} \right)} \times\begin{pmatrix}
  1&0.1&0.2 \\	
  0.1&1.5&0.3\\
  0.2&0.3&1.2
 \end{pmatrix} \otimes \boldsymbol{I}_{10}$ & 2.55\\
$\beta_1$ & 2 & 0.34\\
$\beta_2$ & 0.95 & 0.08\\
\hline
\end{tabular*}
\bigskip
\end{table*}

\begin{figure}[!h]
\centering
\includegraphics[trim=10mm 0mm 0mm -10mm,clip,scale=0.6,angle=270]{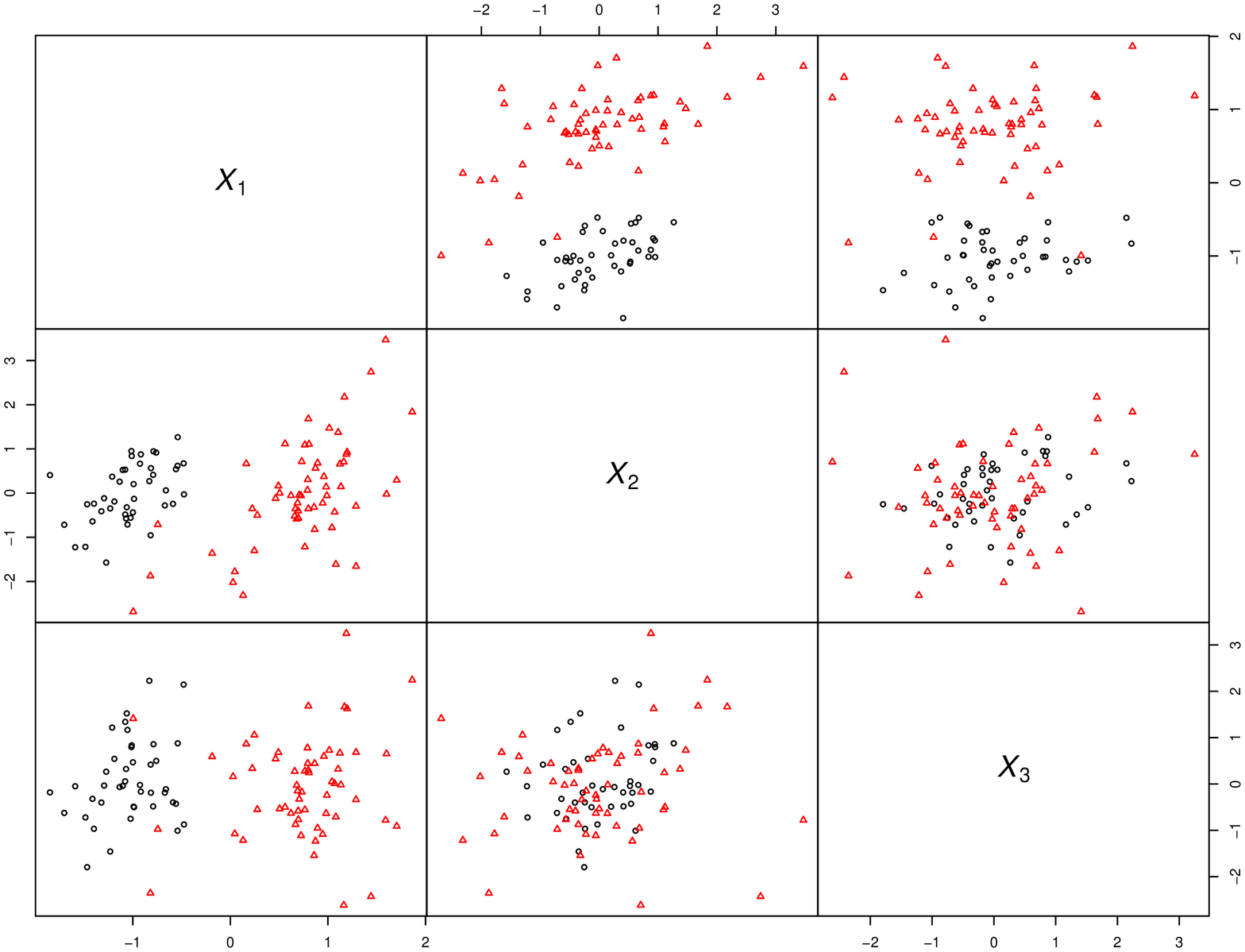}
\caption{Scatter plots showing an example of a generated data set for Simulation~3.}
\label{sim4}
\end{figure}

\paragraph{Simulation 4: Gaussian and $t$-components}
Here, we show that the ePEM family can recover Gaussian and $t$-components favourably when compared to the {\tt mixture} and {\tt teigen} families. A two-component mixture is simulated with 100 observations, where the group sample sizes are sampled from a binomial distribution with success probability $0.4$. The first component is simulated from a 3-dimensional Gaussian distribution with zero mean. The second component is simulated from a 3-dimensional $t$-distribution with mean $(5,0,0)'$ and 5 degrees of freedom. Both components are generated using the same scale matrix: $$\begin{pmatrix}
  1&0.5&0.25 \\	
  0.5&1&0.3\\
  0.25&0.3&1
 \end{pmatrix}.$$ 
 The algorithms are run for $G=1,\ldots,5$. 
The {\tt mixture} family does not perform well over 100 runs. One through five component models are chosen 1, 52, 31, 14, and  2 times, respectively. In contrast, for the ePEM family, a two (three) component model is selected 89 (11) times. On the occasion when a three-component model is selected, the low overall sample size seems to contribute to some observations from the heavy-tailed component being clustered in their own unique group. Similarly, for the {\tt teigen} family, a two (three) component model is selected 88 (12) times. Over the 100 runs, the EEEE (EEEV) model is selected 70 (21) times. Given the generated data, a model with varying $\beta_g$ might be expected from the ePEM family; however, in a few runs, the selected models have heavy tailed components with equal $\beta_g$. This may be due to the small overall sample size and/or the fact that the generated components are not clearly separated. The ARI values for the selected models for the {\tt mixture} family over the 100 runs range from 0 (for the one-component model) to 1, with a median (mean) ARI of 0.94 (0.90). Similarly, the selected models from both the {\tt teigen} and ePEM families yield ARI values ranging between 0.57 and 1, with a median (mean) value of 0.96 (0.94).
A scatter plot showing an example of the generated data is given in Figure \ref{sim4}.

\paragraph{Assessing the impact of outliers}
We follow \cite{mclachlan2000} in assessing the impact of outliers on the clustering performance of the ePEM family as compared to the Gaussian mixture models implemented in the {\tt mixture} package. The {\tt crab} data set, introduced in \cite{campbell1974}, consists of five-dimensional observations on crabs of the genus {\it Leptograpsus} and can be obtained from the {\tt MASS} package \citep{venables2002}. Measurements are recorded on the width of the front lip, the rear width, the length along the middle, the maximum width of the carapace, and the body depth. The subset of blue crabs (50 males and 50 females) is analyzed in \cite{mclachlan2000}, where outliers are introduced, and a Gaussian model with a common covariance matrix as well as a $t$-mixture model with equal scale matrices and equal degrees of freedom are fitted. The outliers are introduced by adding various values to the second variate of the $25^{\text{th}}$ point. We replicate this analysis to investigate the performance of the ePEM models compared to the {\tt mixture} models. Note that the EEEE model from the {\tt teigen} family is also fitted but does not perform well (a minimum of 37 misclassifications; results not shown). This is probably due to different starting values; however, \cite{mclachlan2000} do not provide information on the starting values used for their comparison and we are unable to obtain results similar to theirs. On the original data, Gaussian EEE and MPE EEEE two-component models yield 19 misclassifications each. However, as the value of the constant that is added to the observation of interest is increased or decreased, the MPE component model error rate is much smaller than that of the Gaussian mixture. However, both the Gaussian mixture and MPE approach suffer when the constant by which the value is jittered is extreme.
\begin{table*}
\caption{Comparison of error rates from the Gaussian and MPE mixture models fitted to modified {\tt crabs} data.} \label{tab:outliers}
\begin{tabular*}{1.0\textwidth}{@{\extracolsep{\fill}}lccc}
\toprule
Constant & Gaussian & MPE & $\hat{\beta}$ \\
\midrule
$-15$ &   37  & 35 & 0.43            \\
$-10$ &   40  & 21 & 0.46           \\
$-5$ &   42  & 20 & 0.73           \\
$0$ &   19  & 19 & 0.80           \\
$5$ &   22  & 20 & 0.67           \\
$10$ &   36  & 37 & 0.52           \\
$15$ &   38  & 41 & 0.43           \\
\bottomrule
\end{tabular*}
\bigskip
\bigskip
Entries in the first column are the values added to the second variate of the $25^{\text{th}}$ observation to make it an outlier. Entries in the second and third columns are the number of misclassifications for the Gaussian and MPE mixture models, respectively. Lastly, the $\hat{\beta}$ values are also provided.
\end{table*}

\subsection{Real Data}\label{sec:realdata}
We also test our algorithm's performance on several real benchmark data sets. The {\tt body}, {\tt diabetes}, {\tt female voles}, and {\tt wine} data sets are commonly used for illustration in the model-based clustering literature. We also consider two bioinformatics data sets: the {\tt srbct} data and the {\tt glob} data. The {\tt body} data contain 24 measurements on body dimension, age, weight, and height for 507 individuals (247 men and 260 women), and can be obtained from the {\tt gclus} package \citep{gclus}. The {\tt diabetes} data \citep{reaven1979}, obtained from {\tt mclust}, contains three measurements on 145 subjects from three classes: chemical (36 observations), normal (76 observations), and overt (33 observations). The {\tt female voles} data \citep{airoldi1984} contain seven measurements on age and skull size of 86 females of two species of voles: \emph{Microtus californicus} (41 observations) and \emph{M. Ochrogaster} (45 observations). These data are available as part of the {\tt{Flury}} package \citep{flury1997} in {\sf R}. Lastly, the {\tt wine} data \citep{forina1988} contain 13 measurements on 178 wines of three types (barolo, grignolino, and barbera), and can be obtained from the {\tt gclus} package.

The {\tt srbct} data contain gene expression microarray data from experiments on small round blue cell tumors \citep{khan2001}. A preprocessed version of these data can be obtained from the {\tt plsgenomics} package \citep{plsgenomics}. The 83 samples are known to correspond to four classes, including 29 cases of Ewing sarcoma, 11 cases of Burkitt lymphoma, 18 cases of neuroblastoma, and 25 cases of rhabdomyosarcoma. The {\tt golub} data contain gene expression data from \cite{golub1999} on two forms of acute leukaemia: acute lymphoblastic leukaemia (47 observations) and acute myeloid leukaemia (25 observations). The preprocessed data used in the analysis of \cite{McNi:Murp:Mode:2010} are available at {\tt www.paulmcnicholas.info}. 
Note that methodology proposed herein is not designed for  high-dimensional, low sample size (i.e., large $p$, small $N$) problems --- the development of factor analysis-based extensions of MPE mixture models, along the lines of the mixture of factor analyzers model \citep{ghahramani1996,mcLachlan2000b} and extensions thereof \citep[e.g.,][]{mcnicholas08,Andr:McNi:Exte:2011}, will be a subject of future work. Hence, both of these bioinformatics data sets are further pre-processed to make the clustering problem more suitable for the methodology that is the subject of the present work. A differential expression analysis on the gene expression data is performed using an ANOVA across the known groups. The top ten genes, ranked using the obtained p-values, are selected to represent a potential set of measurements that contain information allowing for identification of the four groups. The three mixture model-based clustering algorithms were then run on these processed data.
The ePEM family is run on the scaled data for $G=1,\ldots,5$. Table~\ref{realdatacomp} compares the performance of the methodologies run on these data; here, the predicted classifications from the selected model (using the BIC) are compared to the true class labels in each case. 
\begin{table*}
\caption{Comparison of three families of mixture models on benchmark data.} \label{realdatacomp}
\begin{tabular*}{1.0\textwidth}{@{\extracolsep{\fill}}lrrr}

  \hline
Data & ePEM & {\tt mixture} & {\tt teigen} \\ 
  \hline
{\tt body} ($p=24$, $G=2$) & 0.94 (2; EEEV) & 0.80 (3; EEE) & 0.80 (3; EEEE)  \\ 
{\tt diabetes} ($p=3$, $G=3$) & 0.66 (3; VVVE) & 0.66 (3; VVV) & 0.67 (3; VVVE)\\ 
{\tt female voles} ($p=7$, $G=2$) & 0.91 (2; EEEV) & 0.91 (2; EEE) & 0.91 (2; EEEE) \\ 
{\tt wine} ($p=13$, $G=3$) & 0.98 (3; EEEV) & 0.68 (4; VVI) & 0.68 (4; VVIE) \\ 
{\tt srbct} ($p=10$, $G=4$) & 0.82 (4; VIIE) & 0.82 (4; VVI) & 0.85 (4; VVIE) \\
{\tt golub} ($p=10$, $G=2$) & 0.84 (2; EEIE) & 0.47 (5; VVE) & 0.74 (2; VVIE)\\
   \hline
\end{tabular*}
\bigskip
\bigskip
Dimensionality and the number of known groups (i.e., classes) are in parenthesis following the name of each data set. For each family of models, the ARI, the number of components, and scale structure for the selected model are given in parenthesis.
\end{table*}

Clearly, the ePEM family performs favourably compared to the {\tt mixture} and {\tt teigen} families. For the {\tt body} data, the selected ePEM model fits a mixture of two heavy tailed components, i.e., $\hat{\bbeta}=(0.57, 0.56)'$, that misclassifies eight cases (4 of each gender). The {\tt teigen} family selects a model with 3 heavy-tailed components (23.43 degrees of freedom each), and the selected {\tt mixture} model also fits three components. 
For the {\tt diabetes} data, the selected models from all three families yield similar classifications, each with a total of 20 misclassifications. The selected ePEM model has $\hat{\beta}=1.07$ in each component, suggesting components that are close to Gaussian. The selected {\tt teigen} model also has relatively high (50.30) degrees of freedom in each component, implying component shapes that are close to Gaussian. 
For the {\tt female voles} data, the selected models from all three families yield the same classification results, each with two misclassifications.
For the {\tt wine} data, both the selected {\tt mixture} and {\tt teigen} models have four components, with 19.15 degrees of freedom in each component for the chosen {\tt teigen} model. However, the selected ePEM model has three components, with $\hat{\bbeta}=(0.62, 0.59, 0.56)'$, and misclassifies only one observation, whereas the selected {\tt mixture} and {\tt teigen} models misclassify 35 and 34 observations, respectively. 

For the {\tt srbct} data, the selected {\tt teigen} model performs slightly better than the selected {\tt mixture} and ePEM models. All selected models fit four components with the selected {\tt teigen}, {\tt mixture}, and ePEM models misclassifying 4, 5, and 5 observations, respectively. The selected {\tt teigen} model has 15.53 degrees of freedom in each component, while the selected ePEM model has $\hat{\beta} = 0.42$ in each component. 

Despite similar outcomes being obtained for the {\tt srbct} data, the results differ greatly for the {\tt golub} data. The selected {\tt mixture} and {\tt teigen} models have five and two components, respectively. A referee asked us to comment on situations where the number of parameters approaches the number of observations. A restriction can be imposed such that only those models are fitted that estimate fewer parameters than the number of observations in the sample. The selected five-component {\tt mixture} model has more parameters than there are observations. Restricting {\tt mixture} to only those models with fewer parameters than the number of observations, a three-component model is selected with an ARI value of 0.76. The selected ePEM model also has two components with $\hat{\beta}=0.28$ in each component, and yields a higher ARI than the selected {\tt teigen} model, which has 5.80 degrees of freedom in each component.

Overall, on these real data sets, the ePEM family outperforms the corresponding family of Gaussian mixtures and performs at least as well as the corresponding mixtures of $t$-distributions. Note that the BIC and the ICL picked the same ePEM model for all real data sets. We also ran these three algorithms on other commonly used data in model-based clustering: the Swiss bank note \citep{flury1997} and the iris \citep{anderson1935, fisher1936} data sets. On these data, the selected models from all three algorithms fit the same number of components and had approximately the same ARI values (results not shown).

\section{Discussion}\label{sec:mpediscussion}

A family of MPE mixture models was proposed based on the density introduced in \cite{gomez1998}. This expanded family of mixture models is introduced with a greatly improved parameter estimation procedure as compared to the techniques proposed previously. This family of mixture models is unique in being able to deal with both lighter and heavier tails than the Gaussian distribution. Mixtures of $t$-distributions can only account for heavier than Gaussian tails and suffers when fitted to lighter tailed data. In such cases, both mixtures of $t$-distributions and mixtures of Gaussian distributions often fit more than the true number of components. Using simulations, we showed that the ePEM family is a good alternative to mixtures of Gaussian and mixtures of Student-$t$ distributions, and that it is able to handle Gaussian, heavy-tailed, and light-tailed components. Moreover, these models also allow for different levels of peakedness of data: from thin to Gaussian to flat. Hence, these models are also well suited for density estimation purposes for a wide range of non-Gaussian data.

Estimation is provided for eight scale structures that can be obtained through the use of eigen-decomposition of the scale matrix. Previously, mixtures of Gaussian and uniform distributions have been fitted to account for outliers \citep{banfield1993, hennig2008, coretto2010}. In our framework, a uniform component can be conveniently approximated by restricting $\beta$ to be high because the power exponential distribution becomes a multivariate generalization of the uniform distribution. This enables greater parsimony than a mixture of Gaussian and uniform distributions when fitted to data with random noise, e.g., on mean-centred data, an EIIE model  requires estimation of only one additional parameter. A mixture of skewed power exponential distributions will be a focus of future work; such a model will be better suited to modelling data with asymmetric clusters. Lastly, note that the ePEM family has heavy fat tails for higher dimensions \citep{liu2008}; therefore, a mixture of power exponential factor analyzers model may be useful for higher-dimensional data with outliers. 

\section*{Acknowledgements}
This work is supported by an Alexander Graham Bell Canada Graduate Scholarship (Dang) and a Discovery Grant (McNicholas) from the Natural Sciences and Engineering Research Council of Canada as well as an Early Researcher Award from the Ontario Ministry of Research and Innovation (McNicholas).

\appendix
\section*{Appendix}

\section{Fixed-point algorithm} \label{sec:Zhangcomparison}
\begin{figure}
\includegraphics[scale=0.25,angle=270]{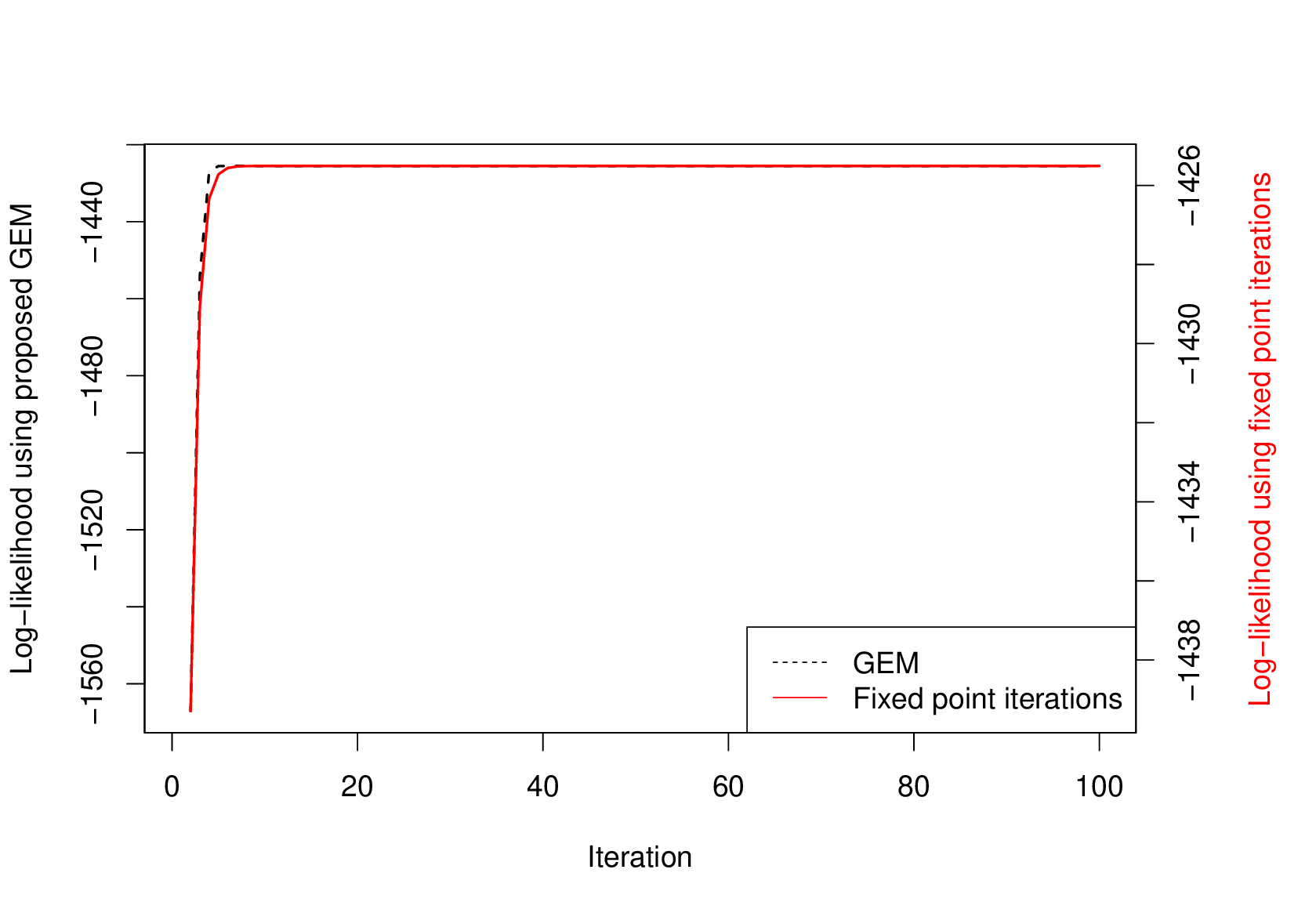}
\includegraphics[scale=0.25,angle=270]{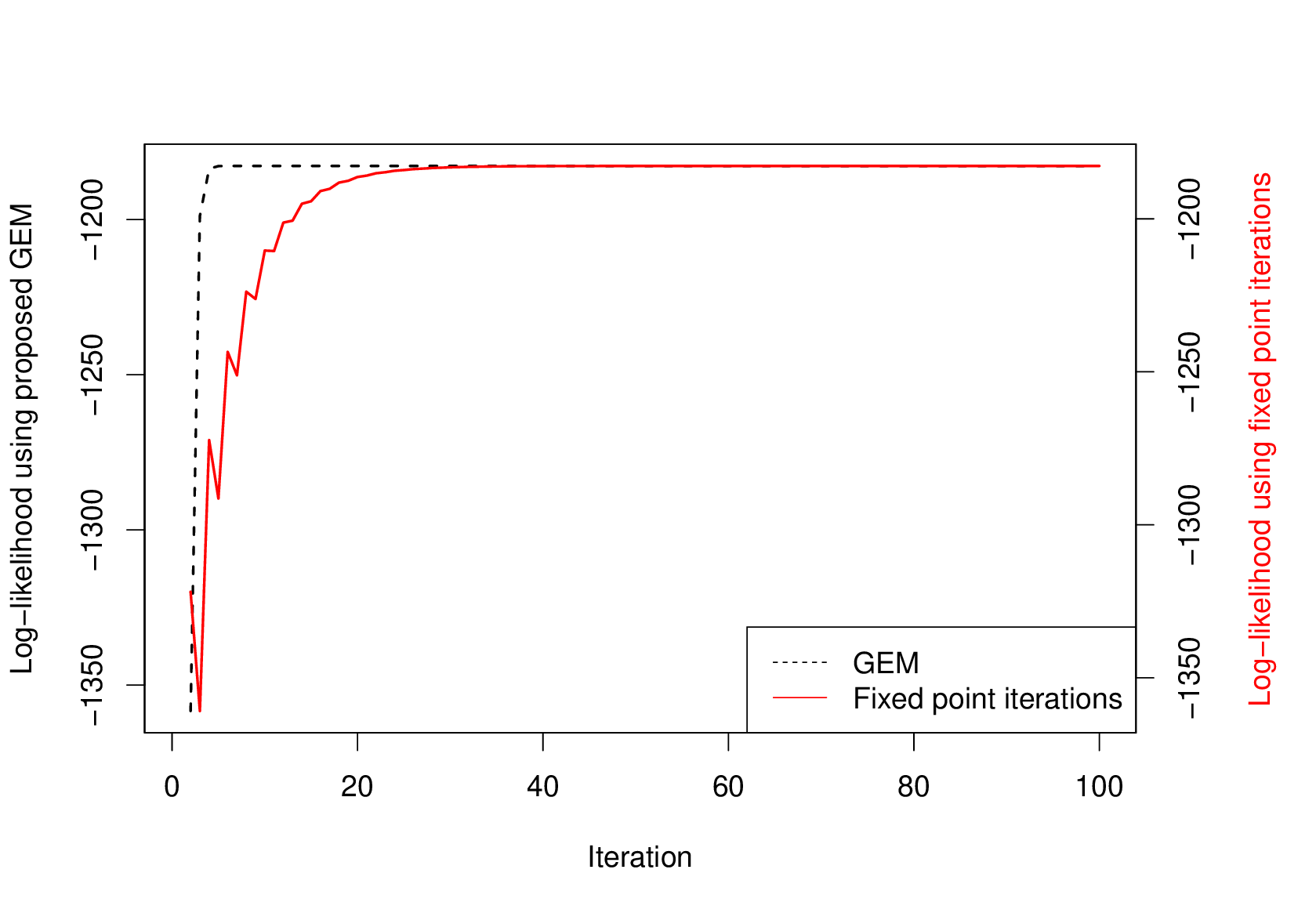}\\
\includegraphics[scale=0.25,angle=270]{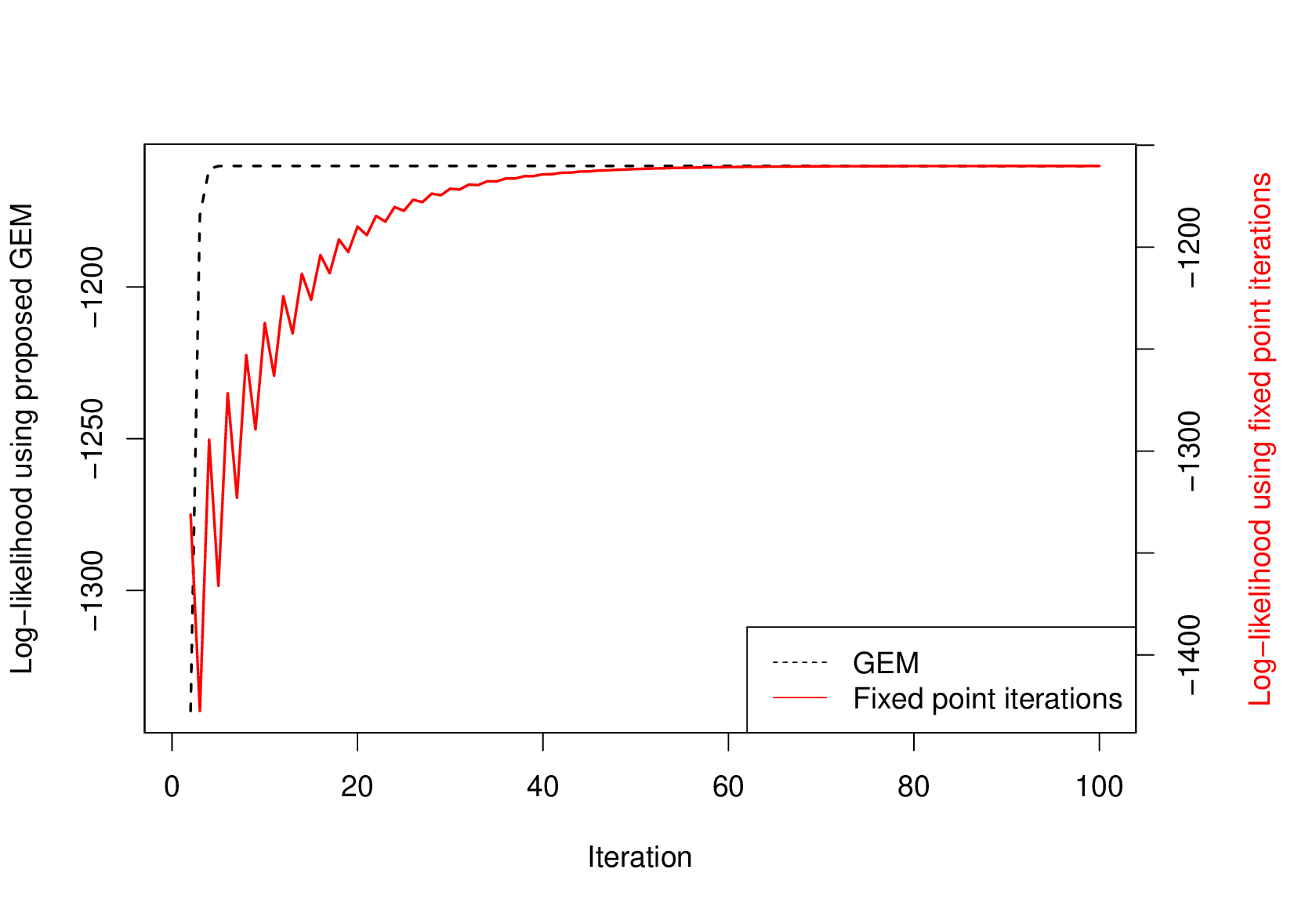}
\includegraphics[scale=0.25,angle=270]{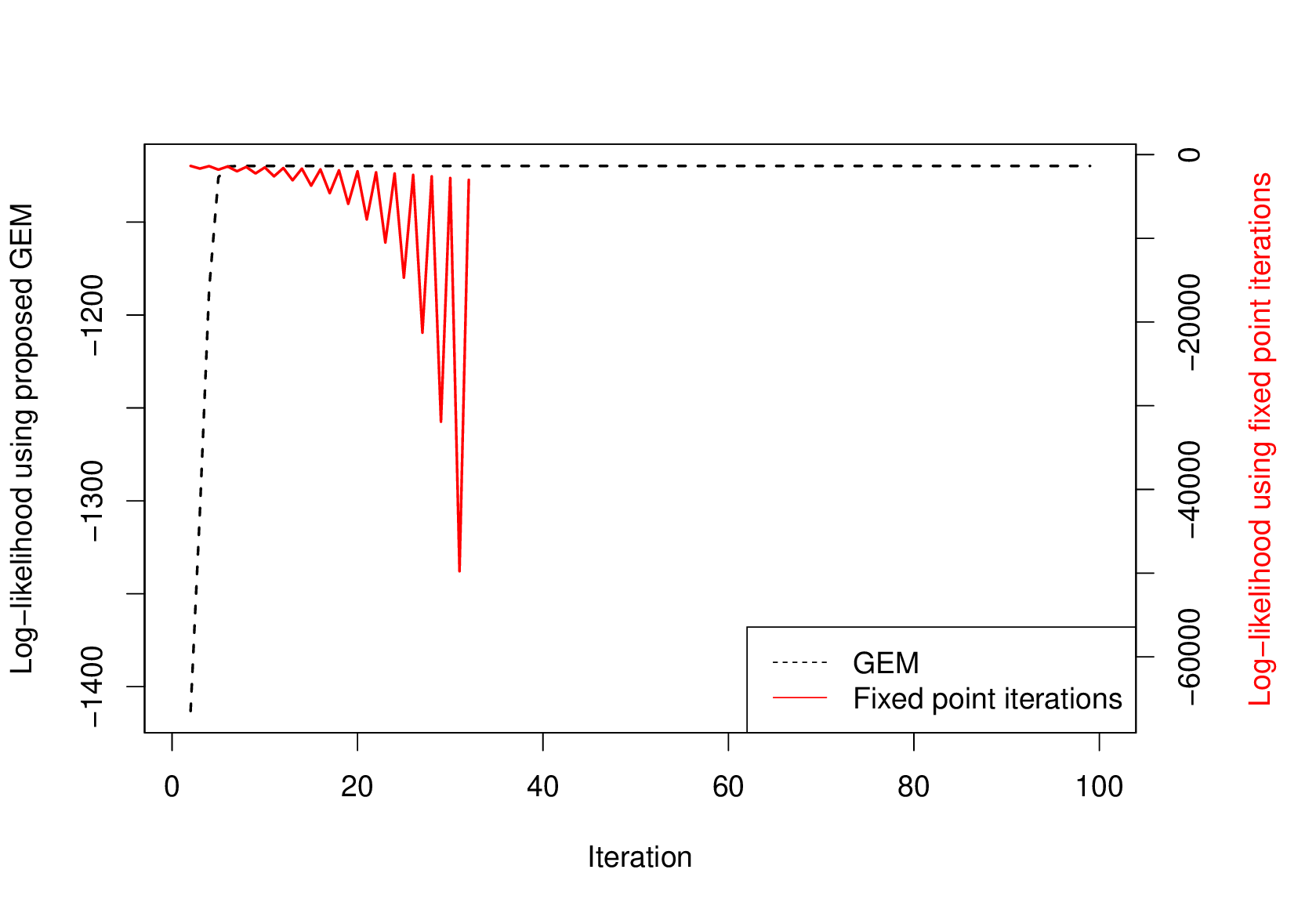}
\caption{Log-likelihood plots for our GEM procedure and fixed point-based estimating algorithms for two-dimensional data. The top-left, top-right, bottom-left and bottom-right panel have $\beta$ equaling 1.5, 1.9, 1.95, and 2.05, respectively.}
\label{compwithfpp2}
\end{figure}

\begin{figure}[!h]
\includegraphics[scale=0.29,angle=270]{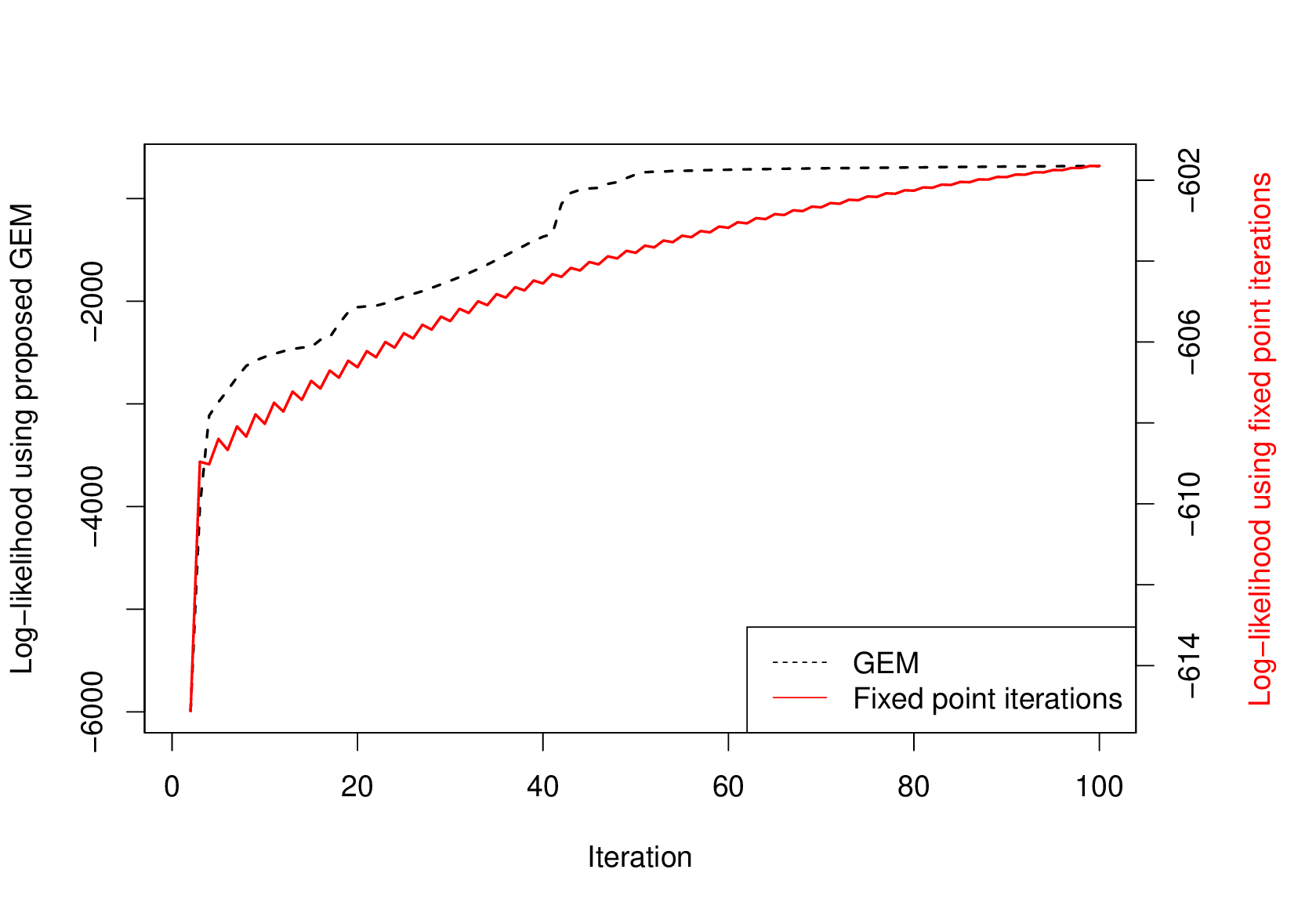}
\includegraphics[scale=0.29,angle=270]{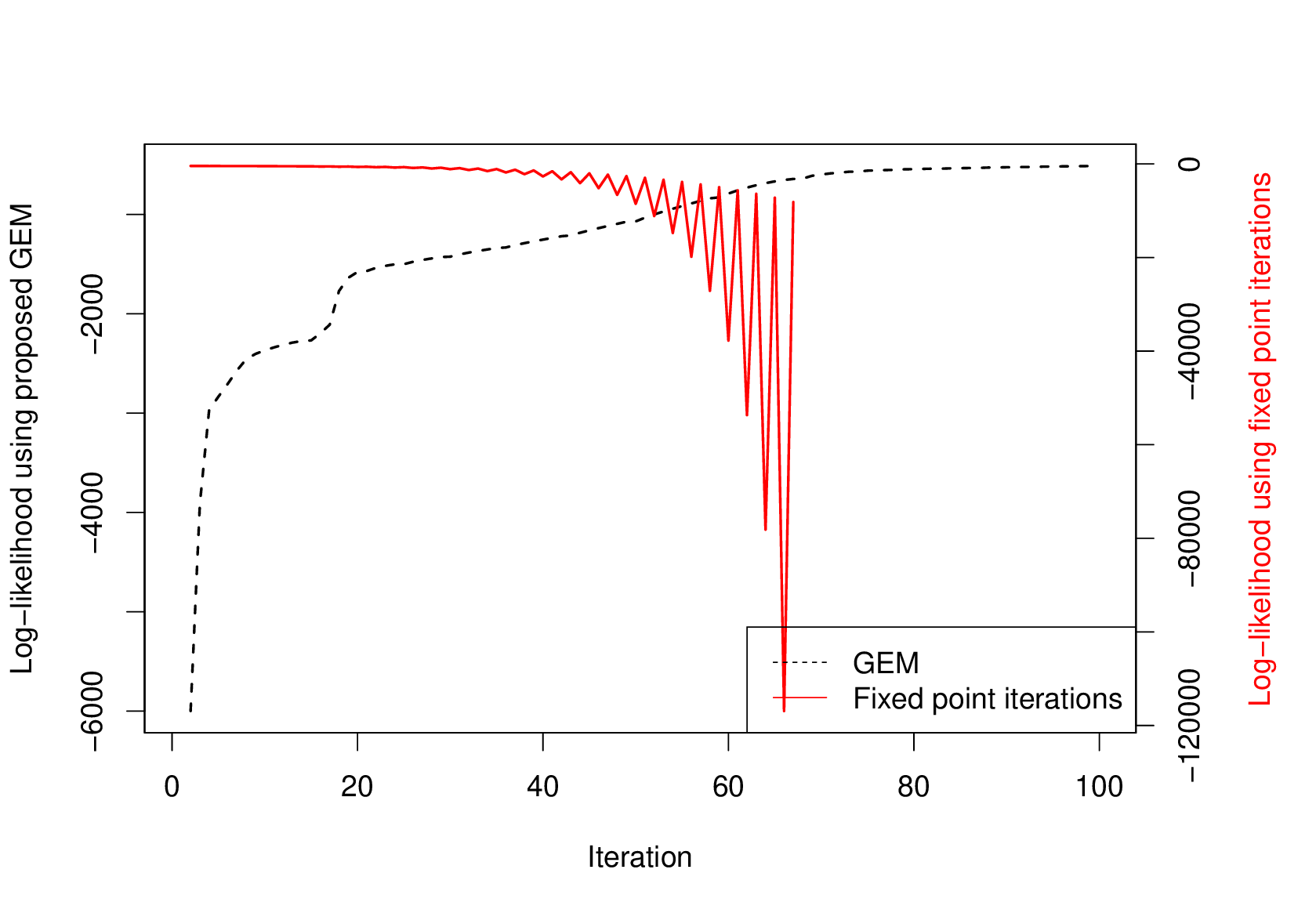}
\caption{Log-likelihood plots for the proposed GEM procedure and fixed point-based estimating algorithms on 10-dimensional data. The left- and right-hand panels have $\beta$ values of 1.99 and 2.05, respectively.}
\label{compwithfpp10}
\end{figure}

\cite{zhang2010} used fixed point iterative estimates for $\bSigma_g$. Note that the MPE density used in \cite{zhang2010} can be obtained by setting $\bSigma=2\bDelta$, $r=2^{\beta^*}$, and $s=\beta^{*}/2$ in \eqref{mpegeneral}, where $\bDelta$ denotes the scale matrix in the parameterization of \cite{zhang2010}. We show that the estimation procedure used in \cite{zhang2010} is valid only for $\beta^{*}\in(0,4)$, where $\beta^{*}$ is defined as in \cite{zhang2010}. A proof for this (see Appendix \ref{sec:fixedpointproof}) applies to $\beta\in(0,2]$, because of the different shape parameterizations, without loss of generality. 
In Figure \ref{compwithfpp2}, we present four comparisons of the trajectory of log-likelihood values between our proposed estimation and using fixed point iterations: for $\beta$ equaling 1.5, 1.9, 1.95, and 2.05, respectively. For all cases, 1000 observations were generated from a 2-dimensional zero-centred power exponential distribution. Only $\bSigma$ is estimated, with the other parameters held constant. For both algorithms, $\bSigma$ is initialized as an identity matrix. Clearly, as $\beta$ approaches 2, the log-likelihood values for the fixed point estimating procedure (red line) oscillate more heavily. This leads to non-monotonicity of the likelihood, complicating the determination of convergence. Moreover, notice that certain values of the log-likelihood for the fixed point are not plotted  for $\beta=2.05$---this is because of numerical errors. Note that each of the subplots in Figure~\ref{compwithfpp2} has two ordinate axes due to different scales of the values from each procedure. We also provide similar plots for $\beta$ equaling 1.99 and 2.05 for a 10-dimensional simulation (Figure \ref{compwithfpp10}). The results are quite similar. We have conducted extensive simulations and, in every case, the log-likelihood values from the fixed point iterations diverge for $\beta>2$. In most cases, the fixed point iterations do not even run. Because this is equivalent to $\beta^{*}>4$, we conclude that the GEM approach is better than using fixed point iterations. Furthermore, note that for $\beta^{*}<2$, the fixed point algorithm for an unconstrained scale matrix converges due to concavity properties (similar to our VVV case for $0<\beta<1$). Note that \cite{zhang2010} only deal with $\beta^*\leq4$ in their work.

\subsection{Fixed point stability} \label{sec:fixedpointproof}
\paragraph{The fixed point algorithm from \cite{zhang2010} diverges for $\beta > 2$}
\begin{proof}
If $\bX$ follows a $p$-dimensional power exponential distribution, the log-likelihood with respect to $\bSigma$ is $$\cL (\bSigma)=\sum_{i=1}^N\sum_{g=1}^G\frac{1}{2} \log|{\bSigma}|^{-1}-\frac{1}{2}  \left[ (\bx_i-\bmu)'\bSigma^{-1}(\bx_i-\bmu) \right]^{\beta}.$$ Then, upon taking the derivative of $\cL (\bSigma)$ with respect to $\bSigma^{-1}$, we can obtain the fixed point update
\begin{equation} \label{fpeqp-dim}
f(\bSigma)=\frac{\beta}{N}\sum_{i=1}^N \left[ (\bx_i-\bmu)'\bSigma^{-1}(\bx_i-\bmu) \right]^{\beta-1} (\bx_i-\bmu) (\bx_i-\bmu)'.
\end{equation}
Now, 
$$\veco (f(\bSigma)) = \frac{\beta}{N} \sum_{i=1}^N \left[ (\bx_i-\bmu)' \bSigma^{-1} (\bx_i-\bmu) \right]^{\beta-1} \veco((\bx_i-\bmu)\otimes (\bx_i-\bmu)).$$ 
Taking the derivative with respect to $\bSigma$, we get the Jacobian 
\begin{align*}
\bJ&=\frac{\beta (1-\beta)}{N}\sum_{i=1}^N \left[ (\bx_i-\bmu)'\bSigma^{-1}(\bx_i-\bmu) \right]^{\beta-2} \\ 
& \qquad\qquad\qquad\qquad\times \veco((\bx_i-\bmu)\otimes (\bx_i-\bmu)) \veco(\bSigma^{-1}(\bx_i-\bmu)(\bx_i-\bmu)'\bSigma^{-1})' \\
&=\frac{\beta (1-\beta)}{N}\sum_{i=1}^N\left\{ \left[ (\bx_i-\bmu)'\bSigma^{-1}(\bx_i-\bmu) \right]^{\beta-2}\right.\\
&\qquad\qquad\qquad\qquad\times \left.\veco((\bx_i-\bmu)\otimes (\bx_i-\bmu)) \left[\veco(\bSigma^{-1}\otimes \bSigma^{-1}) \veco((\bx_i-\bmu) \otimes (\bx_i-\bmu))\right]'\right\}.
\end{align*}
Then, 
\begin{align*}
\tr(\bJ)&=\frac{\beta (1-\beta)}{N}\sum_{i=1}^N \left[ (\bx_i-\bmu)'\bSigma^{-1}(\bx_i-\bmu) \right]^{\beta-2}\\
&\qquad\qquad\qquad\qquad\qquad\times \tr \left\{(\bx_i-\bmu) (\bx_i-\bmu)' \bSigma^{-1}\otimes (\bx_i-\bmu) (\bx_i-\bmu)' \bSigma^{-1}\right\} \\
&=\frac{\beta (1-\beta)}{N}\sum_{i=1}^N \left[ (\bx_i-\bmu)'\bSigma^{-1}(\bx_i-\bmu) \right]^{\beta-2} \\
&\qquad\qquad\qquad\qquad\qquad \times \tr \left\{(\bx_i-\bmu) (\bx_i-\bmu)' \bSigma^{-1}\right\}\tr\left\{(\bx_i-\bmu) (\bx_i-\bmu)' \bSigma^{-1}\right\}\\
&=\tr\left\{(1-\beta)\bSigma^{-1}\frac{\beta}{N}\sum_{i=1}^N \left[ (\bx_i-\bmu)'\bSigma^{-1}(\bx_i-\bmu) \right]^{\beta-1} (\bx_i-\bmu) (\bx_i-\bmu)' \right\}.\\
\end{align*}
Evaluating $\tr(\bJ)$ at $\bSigma=\hat{\bSigma}$, we get $\tr(\bI_p(1-\beta))=(1-\beta)p$. Now, because the matrix norm $||\bB||_p=\tr(\bB^p)^{1/p}$, $||\bJ||_1=\tr(\bJ)=p(1-\beta)$. 
Also, note that because $||J||_1\leq p ||J||_\infty$, and we require $||J||_\infty<1$ for stability ($||J||_\infty=1$ for neutrality), we have
\begin{equation*}
||J||_1\leq p ||J||_\infty<p.
\end{equation*}
Hence, $p(1-\beta)<p$, leading to $0<\beta<2$. Therefore, the solution diverges for $\beta>2$.
\end{proof}

\section{Inference} \label{sec:parupdates}

The likelihood of the MPE mixture model is
\begin{equation*}
L_0(\bTheta|\mathcal{S})=\prod_{i=1}^{N} \sum_{g=1}^{G} \pi_g k_g|\bSigma_g|^{-\frac{1}{2}} \exp\left\{ -\frac{1}{2} \left( \bm_{ig}'\bSigma_g^{-1}\bm_{ig}\right)^{\beta_g} \right\},
\end{equation*}
where $k_g$ is analogous to $k$ in \eqref{mpedef},
and $\bm_{ig}=\bx_i-\bmu_g$.
Note that $\mathcal{S}$ is considered incomplete in the context of the EM algorithm. 
The complete-data are $\mathcal{S}_{\text{c}}=\left\{(\bx_1, \bz_1), \ldots, (\bx_N, \bz_N)\right\}$, where the missing data $\bz_{i}= (z_{i1}, \ldots, z_{iG} )'$ is the component label vector such that $z_{ig}$ = 1 if $\bx_i$ comes from the $g^\text{th}$ population and 0 otherwise. 
The complete-data log-likelihood $\cL_c(\bTheta)=\log{L_c(\bTheta|\mathcal{S}_{\text{c}})}$ can be written as
\begin{equation*}
\cL_c(\bTheta)=\sum_{i=1}^{N} \sum_{g=1}^{G} z_{ig} \log \left[\pi_g k_g
 |\bSigma_g|^{-\frac{1}{2}} \exp\left\{ -\frac{\delta_{ig}(\bx_i)^{\beta_g}}{2} \right\}\right].
\end{equation*}
where $\delta_{ig}(\bx_i):=\delta_i \left(\bx_i|\bmu_g,\bSigma_g \right)=\left (\bx_i-\bmu_g \right)' \bSigma_g^{-1} \left(\bx_g-\bmu_g \right)$.
The E-step involves calculating the expected complete-data log-likelihood, which we denote $\mathcal{Q}$. We need the expected values
\begin{equation} \label{estep}
\tau_{ig}\colonequals  \EE_{\widehat\bTheta} [Z_{ig}|\bx_i] = \frac{\pi_g f \left(\bx_i| \hat\bmu_g, \hat\bSigma_g,\hat\beta_g\right)} {\sum_{j=1}^G \hat\pi_j f \left(\bx_i| \hat\bmu_{j}, \hat\bSigma_{j},\hat\beta_j\right)},
\end{equation} for $i=1,\ldots,N$ and $g=1,\ldots,G$.
The M-step on the $(k+1)$th iteration involves maximization of the expected value of the complete-data log-likelihood with respect to $\bTheta$.
The update for $\hat\pi_{g}$ is $$\hat{\pi}_{g} = {n_g}/{N},$$ where $n_g=\sum_{i=1}^N\tau_{ig}$.

However, the updates for $\hat{\beta}_g$, $\hat{\bmu}_{g}$, and $\hat{\bSigma}_{g}$ are not available in closed form. 
A Newton-Raphson update is used to find the update for $\hat{\bmu}_g$, and we need the following:
\begin{align} 
\frac{\partial \mathcal{Q}}{\partial \bmu_g} =& \hat{\beta}_g\sum_{i=1}^{N} \tau_{ig} \delta_{ig}(\bx_i)^{\hat{\beta}_g-1} \hat{\bSigma}_g^{-1} \bm_{ig} \label{muNR1}\\
\frac{\partial^2 \mathcal{Q}}{\partial \bmu_g\bmu_g'} =& \hat{\beta}_g\sum_{i=1}^{N} \tau_{ig} \left[-\delta_{ig}(\bx_i)^{\hat{\beta}_g-1} \hat{\bSigma}_g^{-1} 
 + (\hat{\beta}_g-1)\delta_{ig}(\bx_i)^{\hat{\beta}_g-2}\hat{\bSigma}_g^{-1} \bm_{ig}\big(-2\hat{\bSigma}_g^{-1} \bm_{ig}\big)'\right],\label{muNR2}
\end{align}
where $\delta_{ig}(\bx_i):=
\left (\bx_i-\hat{\bmu}_g \right)'\hat{\bSigma}_g^{-1} \left(\bx_i-\hat{\bmu}_g \right)$ and $\bm_{ig}=\bx_i-\hat{\bmu}_g$. An update for $\hat{\beta}_g$ can be obtained by solving the equation
\begin{align} \label{betasolve}
\frac{p n_g}{\left(\hat\beta_g^{\new}\right)^2} &\psi\left(1+\frac{p}{2\hat\beta_g^{\new}}\right)+\frac{p n_g \log 2}{\left(\hat\beta_g^{\new}\right)^2} -\sum_{i=1}^{N}\tau_{ig}[\log{\delta_{ig}(\bx_i)}]\left(\delta_{ig}(\bx_i)\right)^{\hat\beta_g^{\new}}=0
\end{align}
for $\hat\beta_g^{\new}$, where $\psi(\cdot)$ is the digamma function. Alternatively, a Newton-Raphson method might be implemented using the following:
\begin{align}
\frac{\partial \mathcal{Q}}{\partial \beta_g} =&  \frac{p n_g}{2 \hat\beta_g^{2}} \psi\left(1+\frac{p}{2\hat\beta_g}\right) +\frac{pn_g\log{2}}{2 \hat\beta_g^{2}} - \sum_{i=1}^N \frac{\tau_{ig}}{2} \delta_{ig}(\bx_i)^{\hat\beta_g} \log{\delta_{ig}(\bx_i)} \label{betaNR1}\\
\frac{\partial^2 \mathcal{Q}}{\partial \beta_g^2} =&  \frac{-p n_g}{\hat\beta_g^{3}} \psi\left(1+\frac{p}{2\hat\beta_g}\right) -  \frac{p^2 n_g}{4 \hat\beta_g^{4}} \psi_1\left(1+\frac{p}{2\hat\beta_g}\right) \label{betaNR2} - \frac{pn_g\log{2}}{\hat\beta_g^{3}} -\sum_{i=1}^N \frac{\tau_{ig}}{2} \delta_{ig}(\bx_i)^{\hat\beta_g} \left[\log{\delta_{ig}(\bx_i)}\right]^2,
\end{align}
where $\psi_1(\cdot)$ is the trigamma function. Uptates for $\hat\beta_g$ when it is constrained to be equal between groups can be obtained similarly. 
Because the update for $\hat\bSigma_{g}$ is not available in closed form, we rely on convexity properties. For the updates for the EEI, VVI, EEE, EEV, VVE, and VVV scale matrices, we utilize a minorization-maximization step. Because of the properties of a minorization-maximization algorithm, this step increases the expected value of the complete-data log-likelihood at every iteration, thus making the estimation algorithm a generalized EM (GEM) algorithm. In addition, for the EEE, EEV, VVE, and VVV scale matrices, we utilize an accelerated line search method on the orthogonal Stiefel manifold \citep[cf.][]{absil2009, browne2014}. An MM algorithm can be constructed by using the convexity of the objective function---a surrogate minorizing function is employed that is maximized. Note that the surrogate function constructed in the E-step in an EM algorithm is, up to a constant, a minorizing function \citep{hunter2004}. For the EII, VII, EEI, VVI, EEE, EEV, VVE, and VVV scale structures (as listed in Table \ref{tab:models}), the updates are discussed below. The pseudo-code for the estimation of parameters is:
\begin{enumerate}
\item Initialize $\hat\beta_g$, $\hat\bmu_g$, $\hat\bSigma_g$. Compute \eqref{estep}.
\item Update $\hat\beta_g$ using either \eqref{betasolve} or \eqref{betaNR1} and \eqref{betaNR2}; or \eqref{equalbetasolve} or \eqref{equalbetaNR1} and \eqref{equalbetaNR2}, depending on whether $\beta_g$ is unconstrained between groups or not. \label{betapseudoupdate}
\item CM step 1: Update $\hat\bmu_g$ using \eqref{muNR1} and \eqref{muNR2}.
\item CM step 2: Update $\hat\bSigma_g$ depending on the scale structure.
\item Check for convergence. If not converged, go back to Step \ref{betapseudoupdate}.
\end{enumerate}

\subsection{Shape parameter constrained between groups} \label{sec:equalbetasolve}
When $\beta_g$ is constrained to be equal between groups, the update for $\hat{\beta}$ can be obtained by solving the equation
\begin{align} \label{equalbetasolve}
\frac{p N}{\left(\hat\beta^{\new}\right)^{2}} \psi\left(1+\frac{p}{2\hat\beta^{\new}}\right)+\frac{p N \log 2}{\left(\hat\beta^{\new}\right)^{2}} - \sum_{g=1}^{G}\sum_{i=1}^{N}\tau_{ig}[\log{\delta_{ig}(\bx_i)}]\delta_{ig}(\bx_i)^{\hat\beta^{\new}}=0
\end{align}
for $\hat\beta^{\new}$. Alternatively, a Newton-Raphson method might be implemented using the following:
\begin{align} 
\frac{\partial \mathcal{Q}}{\partial \beta} =&  \frac{p N}{2\hat\beta^{2}} \psi\left(1+\frac{p}{2\hat\beta} \right) + \frac{p N \log 2}{2\hat\beta^{2}} - \sum_{g=1}^{G} \sum_{i=1}^{N} \frac{\tau_{ig}}{2}[\log{\delta_{ig}(\bx_i)}]\left(\delta_{ig}(\bx_i)\right)^{\hat\beta} \label{equalbetaNR1}\\
\frac{\partial^2 \mathcal{Q}}{\partial \beta^2} =&  \frac{-p N}{\hat\beta^{3}} \psi\left(1+\frac{p}{2\hat\beta}\right) -  \frac{p^2 N}{4 \hat\beta^{4}} \psi_1\left(1+\frac{p}{2\hat\beta}\right) - \frac{pN\log{2}}{\hat\beta^{3}} \label{equalbetaNR2}
 -\sum_{g=1}^G \sum_{i=1}^N \frac{\tau_{ig}}{2} \left[\log{\delta_{ig}(\bx_i)}\right]^2 \delta_{ig}(\bx_i)^{\hat\beta} .
\end{align}

\subsection{Scale structure VVV} \label{sec:VVV}
Here, details are provided on estimation of the unconstrained scale matrix (VVV structure). On ignoring terms not involving $\bSigma_g$, we have
\begin{equation*}
\mathcal{Q}(\bSigma_g)=\sum_{i=1}^N\sum_{g=1}^G\frac{\tau_{ig}}{2} \log|{\bSigma_g}|^{-1}-\frac{\tau_{ig}}{2}  \left( \bm_{ig}'\bSigma_g^{-1}\bm_{ig} \right)^{\beta_g}.
\end{equation*}
The updates differ based on the value of $\hat\beta_g^{\new}$. Denote $\bM_{ig}^{\new}=\bm_{ig} \bm'_{ig}$, where $\bm_{ig}=\bx_i-\hat{\bmu}^{\new}_g$. 

$\hat\beta_g^{\new}\in(0,1)$:
Here, we borrow from the minorization-maximization framework for estimation. Note that $\tr\left\{\bSigma_g^{-1}\bM_{ig}\right\}^{\beta_g}$ is concave for $\beta_g\in(0,1)$, where $\tr(\cdot)$ refers to the trace.  A surrogate function for $\tr\left\{\bSigma_g^{-1}\bM_{ig}^{\new}\right\}^{\beta_g^{\new}}$ can be constructed using the supporting hyperplane inequality:
\begin{align*}
\tr\left\{\bSigma_g^{-1}\bM_{ig}^{\new}\right\}^{\beta_g^{\new}} \leq \tr\left\{\hat\bSigma_g^{-1}\bM_{ig}^{\new}\right\}^{\beta_g^{\new}} +&\beta_g^{\new} \tr\left\{\hat\bSigma_g^{-1}\bM_{ig}^{\new}\right\}^{\beta_g^{\new}-1}\\
& \times \left[\tr\left\{\bSigma_g^{-1}\bM_{ig}^{\new}\right\}-\tr\left\{\hat\bSigma_g^{-1}\bM_{ig}^{\new}\right\}\right].
\end{align*}
Then, the following is maximized:
\begin{align*} 
\sum_{i=1}^N\sum_{g=1}^G &-\frac{\tau_{ig}}{2} \log|{\bSigma_g}|^{-1}+\frac{\tau_{ig}}{2}  \left[\tr\left\{\hat\bSigma_g^{-1}\bM_{ig}^{\new}\right\}^{\hat\beta_g^{\new}}  \right. \\
& \left. +\hat\beta_g^{\new}\tr\left\{\hat\bSigma_g^{-1}\bM_{ig}^{\new}\right\}^{\hat\beta_g^{\new}-1} \times \left(\tr\left\{\bSigma_g^{-1}\bM_{ig}^{\new}\right\} - \tr\left\{\hat\bSigma_g^{-1}\bM_{ig}^{\new}\right\}\right)\right],
\end{align*}
leading to the update
\begin{equation}\label{VVV01}
\hat{\bSigma}_g^{\new}=\frac{\hat{\beta}_g^{\new}}{n_g} \sum_{i=1}^N \tau_{ig}\tr\left\{\hat\bSigma_g^{-1}\bM_{ig}^{\new}\right\}^{\hat{\beta}_g^{\new}-1} \bM_{ig}^{\new}.
\end{equation}

$\hat\beta_g^{\new}\in[1,\infty)$: Using the Jordan decomposition, $\bSigma_g^{-1}=\bD_g\bA_g^{-1}\bD_g',$ where $\bD_g$ is an orthonormal matrix and $\bA_g$ is a diagonal matrix of eigenvalues. Now, let $\bSigma_g^{-1}=\bD_g\bLambda_g^{1/\beta_g^{\new}}\bD_g'$ where $\bLambda_g^{-1/\beta_g^{\new}}=\bA_g$. We obtain updates for both $\bA_g^{\new}$ and $\bD_g^{\new}$. 

It follows that  $$\tr\left\{\bm'_{ig}\hat{\bD}_g\bLambda_g^{1/\beta_g^{\new}}\hat{\bD}'_g \bm_{ig}\right\}^{\beta_g^{\new}} = \tr\left\{\bv'_{ig}\bLambda_g^{1/\beta_g^{\new}}\bv_{ig} \right\}^{\beta_g^{\new}},$$ where $\bv_{ig}=\hat{\bD}'_g \bm_{ig}$. 
Then, for $i=1,\ldots,N$, $$f(\blambda_g)=\tr\left\{\bv'_{ig}\bLambda_g^{1/\beta_g^{\new}}\bv_{ig} \right\}^{\beta_g^{\new}}=\left(\sum_{h=1}^p \lambda_{gh}^{1/\beta_g^{\new}} v_{igh}^2\right)^{\beta_g^{\new}},$$ where $\bLambda_g=\mbox{diag}(\lambda_{g1},\ldots,\lambda_{gp})$.
This function is concave with respect to the eigenvalues $\blambda_g=\{\lambda_{g1},\ldots,\lambda_{gp}\}$ (cf.\ weighted $p$-norm). A surrogate function is constructed using $$f(\blambda_g) \leq f(\hat\blambda_g) + (\nabla f(\hat\blambda_g))'(\blambda_g-\hat\blambda_g),$$ i.e.,
\begin{align*}
f(\blambda_g) \leq& \left[\sum_{h=1}^p(\hat\lambda_{gh})^{{1}/{\beta_g^{\new}}} v_{igh}^2\right]^{\beta_g^{\new}} + \left[\sum_{h=1}^p (\hat\lambda_{gh})^{{1}/{\beta_g^{\new}}} v_{igh}^2\right]^{\beta_g^{\new}-1}\\& \times \left[\left(v_{ig1}^2 \lambda_{g1}^{{1}/{\beta_g^{\new}}-1},\ldots,v_{igp}^2 \lambda_{gp}^{{1}/{\beta_g^{\new}}-1}\right) \left((\lambda_{g1}-\hat{\lambda}_{g1}) ,\ldots,(\lambda_{gp}-\hat{\lambda}_{gp}) \right)'\right].
\end{align*}
This can be simplified to
\begin{align*}
f(\blambda_g) \leq& \tr\left\{\hat\blambda_g^{{1}/{\beta_g^{\new}}}\bV_{ig} \right\}^{\beta_g^{\new}} + \tr\left\{\bv'_{ig}\hat\blambda_g^{{1}/{2\beta_g^{\new}}}(\hat\blambda_g)^{{1}/{2\beta_g^{\new}}}\bv_{ig} \right\}^{\beta_g^{\new}-1}\\& \times \left( \bv'_{ig}\hat\blambda_g^{{1}/{2\beta_g^{\new}}-{1}/{2}}\bLambda_g \hat\blambda_g^{{1}/{2\beta_g^{\new}}-{1}/{2}} \bv_{ig}-\tr\left\{\bv'_{ig}\hat\blambda_g^{{1}/{2\beta_g^{\new}}}\hat\blambda_g^{{1}/{2\beta_g^{\new}}}\bv_{ig} \right\}\right),
\end{align*}
where $\bV_{ig}=\bv_{ig}\bv'_{ig}$. Now, let $\bW_{ig}=\bw_{ig}\bw'_{ig}$, where $\bw_{ig} = \bv'_{ig} \hat\blambda_g^{1/2\beta_g^{\new}}$. Also, note that here, $\bw_{ig}\bw'_{ig}(\bw'_{ig}\bw_{ig})^{\beta_g^{\new}-1}=(\bw_{ig}\bw'_{ig})^{\beta_g^{\new}}=\bW_{ig}^{\beta_g^{\new}}$. Now, 
\begin{equation*}
f(\blambda_g) \leq \tr\left\{\hat\blambda_g^{{1}/{\beta_g^{\new}}}\bV_{ig} \right\}^{\beta_g^{\new}} + 
\left(\tr\left\{\bLambda_g \hat\blambda_g^{-{1}/{2}} \bW_{ig}^{\beta_{g}^{\new}}\hat\blambda_g^{-{1}/{2}}\right\} -\tr\left\{\bW_{ig}^{\beta_g^{\new}}\right\}\right).
\end{equation*}
Then, the following is maximized:
\begin{align*}
\sum_{i=1}^N\sum_{g=1}^G -\frac{\tau_{ig}}{2\hat\beta_g^{\new}} \log|{\bLambda_g}|+&\frac{\tau_{ig}}{2}  \left[\tr\left\{\hat\blambda_g^{{1}/{\hat\beta_g^{\new}}}\bV_{ig} \right\}^{\hat\beta_g^{\new}} + \right. \\
&\left. \left(\tr \left\{ \bLambda_g \hat\blambda_g^{-{1}/{2}} \bW_{ig}^{\beta_{g}^{\new}} \hat\blambda_g^{-{1}/{2}} \right\} -\tr\left\{\bW_{ig}^{\hat\beta_g^{\new}}\right\}\right)\right].
\end{align*}
On taking the derivative with respect to $\bLambda_g$, it can be shown that the update for $\hat{\bA}_g$ is
\begin{equation} \label{VVV1infdiag}
\hat{\bA}_g^{\new}=\left(\frac{\hat{\beta}_g^{\new}}{n_g}\sum_{i=1}^N \tau_{ig}\hat\bA_g^{{\hat{\beta}_g^{\new}}/{2} }\hat\bW_{ig}^{\hat{\beta}_{g}^{\new}} \hat\bA_g^{{\hat{\beta}_g^{\new}}/{2} }\right)^{1/\hat\beta_g^{\new}},
\end{equation}
where $$\hat\bW_{ig}=\hat\bA_g^{-{1}/{2}}\hat\bV_{ig}\hat\bA_g^{-{1}/{2}},$$
$\hat\bV_{ig}=\hat\bv_{ig}\hat\bv_{ig}'$ and $\bv_{ig}=\hat\bD_g'\bm_{ig}$.
 
Regarding the update for $\hat{\bD}_g$ (this is the same as $\bGamma_g$ in Table \ref{tab:models}), an orthonormal matrix, we use an accelerated line search for optimization on the orthogonal Stiefel manifold as employed by \cite{browne2014}. For minimizing a function of an orthonormal matrix, the search space is the orthogonal Stiefel manifold equal to the set of all orthonormal matrices $\mathcal{M}=\{\bX \in \mathbb{R}^{p\times p}: \bX' \bX=\bI_p\}$. The idea behind the line search method is to move along a specific search direction in the tangent space until the objective function is reasonably decreased \citep{browne2014}. Let $\bQ_{g}= \sum_{i=1}^N \tau_{ig}^{{1}/{\beta_g^{\new}}}\bM_{ig}^{\new}$. The objective function that needs to be minimized is 
$$f(\bD_g)= \sum_{g=1}^G \tr \{\bQ_g\bD_g\left(\hat{\bA}_g^{\new}\right)^{-1}\bD_g'\}^{\hat\beta_g^{\new}},$$
with an unconstrained gradient $$ \bar{\text{grad}f(\bD_g)}=2\hat\beta_g^{\new}\left(\bQ_g{\bD}_g\left(\hat{\bA}_g^{\new}\right)^{-1}{\bD}'_g\right)^{(\hat\beta_g^{\new}-1)}\bQ_g{\bD}_g\left(\hat{\bA}_g^{\new}\right)^{-1}=\mathbf{R}_g .$$ As shown by \cite{browne2014}, the direction of the steepest descent while in $T_{\bX}\mathcal{M}$ (the tangent space of $\bX$) at the position $\bX$ is
$\text{grad} f(\bX)  = \mathbf{P}_{\bX} \left(\bar{\text{grad}f(\bX)}\right),$ where $$\mathbf{P}_{\bX} \left(\bZ\right)=\bZ-\bX \frac{(\bX'\bZ+\bZ'\bX)}{2}$$ is the orthogonal projection $\mathbf{P}_{\bX}$ of a matrix $\bZ$ onto $T_{\bX}\mathcal{M}$. Hence, we get $$\text{grad} f(\bD_g) =\mathbf{R}_g -\frac{1}{2}{\bD}_g\mathbf{R}_g'{\bD}_g-\frac{1}{2}{\bD}_g{\bD}_g'\mathbf{R}_g.$$
In order to obtain convergence, the step size $t^*$ is taken to be the Armijo step size (which guarantees convergence) and $\hat\bD_g$ is updated as
\begin{equation} 
\hat{\bD}_g^{\new}=\mathbf{R}_{\bX}\left[ -t^*_k \times \text{grad} f(\hat\bD_g) \right], \label{VVV1inforth}
\end{equation} 
where $\mathbf{R}_{\bX}$ is a retraction $\mathbf{R}$ at $\bX$. A retraction --- a smooth mapping from the tangent space to the manifold --- allows for searching along a curve in the manifold (while moving in the direction of the tangent vector). As in \cite{browne2014}, the QR decomposition-based retraction is used herein. See \cite{browne2014} for details on the retraction and the Armijo step size.

\subsection{Scale structure VVI}
There are two solutions depending on the current estimate of $\beta_g$. Denote $\bM_{ig}^{\new}=\bm_{ig} \bm'_{ig}$, where $\bm_{ig}=\bx_i-\hat{\bmu}^{\new}_g$.\\

$\hat\beta_g^{\new}\in(0,1)$: Using the Jordan decomposition, we can write $\bSigma_{g}^{-1}=\bA_g^{-1}$, because $\bD_g$ is an identity matrix. 
Recall that the VVI scale structure refers to a diagonal constraint such that  $\bSigma_{g}^{-1}=\bA_g^{-1}=\bLambda_g=\mbox{diag}(\lambda_{g1},\ldots,\lambda_{gp})$, where $\mbox{diag}(\cdot)$ denotes a diagonal matrix. Note that
$\tr\left\{\bLambda_g\bm_{ig}\bm_{ig}'\right)^{\hat{\beta}^{\new}_g}$ can be written as $\left(\sum_{h=1}^p \left(x_{ih}-\mu_{gh}\right)^2 \lambda_{gh}\right)^{\hat{\beta}^{\new}_g}$. This is a concave function with respect to the eigenvalues of the diagonal matrix. 
Then, a surrogate function can be constructed:
\begin{align*}
\tr\left\{\bLambda_g\bM_{ig}^{\new}\right\}^{\hat{\beta}_g^{\new}} \leq & \tr\left\{\hat\bLambda_g\bM_{ig}^{\new}\right\}^{\hat{\beta}_g^{\new}} +\hat{\beta}_g^{\new}\tr\left\{\bm'_{ig}\hat\bLambda_g\bm_{ig}\right\}^{\hat{\beta}_g^{\new}-1} \\
& \times \left( \left(m_{ig1}^2,\ldots,m_{igp}^2\right) \left[ \left(\lambda_{g1}-\hat{\lambda}_{g1}\right) ,\ldots,\left(\lambda_{gp}-\hat{\lambda}_{gp}\right) \right]'\right).
\end{align*}
This leads to 
\begin{align*}
\tr\left\{\bLambda_g\bM_{ig}^{\new}\right\}^{\hat{\beta}_g^{\new}} \leq \tr\left\{\hat\bLambda_g\bM_{ig}^{\new}\right\}^{\hat{\beta}_g^{\new}} +\hat{\beta}_g^{\new} & \tr\left\{\bm'_{ig}\hat\bLambda_g\bm_{ig}\right\}^{\hat{\beta}_g^{\new}-1} \\ 
& \times \left[ \tr\left\{\bLambda_g \bM_{ig}^{\new}\right\}-\tr\left\{\hat\bLambda_g\bM_{ig}^{\new} \right\}\right].
\end{align*}
Then, we maximize:
\begin{align*}
\sum_{i=1}^N\sum_{g=1}^G-\frac{\tau_{ig}}{2} \log|{\bLambda_g}| + & \frac{\tau_{ig}}{2}  \left[\tr\left\{\hat\bLambda_g\bM_{ig}^{\new}\right\}^{\hat\beta_g^{\new}} +\hat\beta_g^{\new}\tr\left\{\hat\bLambda_g\bM_{ig}^{\new}\right\}^{\hat\beta_g^{\new}-1} \right. \\
& \left. \times \left(\tr\left\{\bLambda_g\bM_{ig}^{\new}\right\}-\tr\left\{\hat\bLambda_g\bM_{ig}^{\new}\right\}\right)\right].
\end{align*}
On taking the derivative with respect to $\bLambda_g$, we obtain the update
\begin{equation}  \label{VVI01}
\hat{\bSigma}_g^{\new}=\frac{\hat{\beta}_g^{\new}}{n_g} \sum_{i=1}^N \tau_{ig}\tr\left\{\hat\bSigma_g^{-1}\bM_{ig}^{\new}\right\}^{\hat{\beta}_g^{\new}-1} \bM_{ig}^{\new}.
\end{equation}\\

$\hat\beta_g^{\new}\in[1,\infty)$:  Using the Jordan decomposition, we can write $\bSigma_{g}^{-1}=\bA_g^{-1}$, because $\bD_g$ is an identity matrix. Let $\bSigma_g^{-1}=\bA_g^{-1}=\bLambda_g^{1/\hat\beta_g^{\new}}$. Proceeding in a similar fashion to the $\bA_g$ update in the VVV ($\hat\beta_g^{\new}\in(1,\infty)$) case, we can get the update
\begin{equation} \label{VVI1inf}
\hat{\bSigma}_g^{\new}=\left(\frac{\hat{\beta}_g^{\new}}{n_g}\sum_{i=1}^N \tau_{ig}\hat\bSigma_g^{\frac{\hat{\beta}_g^{\new}}{2}} \bW_{ig}^{\hat{\beta}_{g}^{\new}} \hat\bSigma_g^{\frac{\hat{\beta}_g^{\new}}{2}}\right)^{1/\hat{\beta}_g^{\new}},
\end{equation}
where $\bW_{ig}=\hat\bSigma_g^{-\frac{1}{2}} \bM_{ig}^{\new} \hat\bSigma_g^{-\frac{1}{2}}$.

\subsection{Scale structure VVE}
There are two solutions depending on the current estimate of $\hat\beta_g$. We use similar ideas as in the VVV and VVI cases. Denote $\bM_{ig}^{\new}=\bm_{ig} \bm'_{ig}$, where $\bm_{ig}=\bx_i-\hat{\bmu}^{\new}_g$.\\

$\hat\beta_g^{\new}\in(0,1)$: Using the Jordan decomposition, we write $\bSigma_g^{-1}=\bD\bA_g^{-1}\bD'$. Now, let $\bSigma_g^{-1}=\bD\bLambda_g\bD'$, where $\bLambda_g^{-1}=\bA_g$. Proceeding as before, a surrogate function can be constructed such that
\begin{align*}
\tr\left\{\bLambda_g\bV_{ig}\right)^{\hat{\beta}_g^{\new}} \leq & \tr\left\{\hat\bLambda_g\bV_{ig}\right\}^{\hat{\beta}_g^{\new}} +\hat{\beta}_g^{\new}\tr\left\{\bv'_{ig}\hat\bLambda_g\bv_{ig}\right\}^{\hat{\beta}_g^{\new}-1} \\
& \times \left[ \tr\left\{\bLambda_g \bV_{ig}\right\}-\tr\left\{\hat\bLambda_g\bV_{ig} \right\}\right],
\end{align*}
where $\bv_{ig}=\hat{\bD} \bm_{ig}$ and $\bV_{ig} = \bv_{ig} \bv_{ig}$. Then, we maximize:
\begin{align*}
\sum_{i=1}^N\sum_{g=1}^G-\frac{\tau_{ig}}{2} \log|{\bLambda_g}|+&\frac{\tau_{ig}}{2}  \left[\tr\left\{\hat\bLambda_g\bV_{ig}\right\}^{\hat\beta_g^{\new}} +\hat\beta_g^{\new}\tr\left\{\hat\bLambda_g\bV_{ig}\right\}^{\hat\beta_g^{\new}-1} \right. \\
& \left. \times \left(\tr\left\{\bLambda_g\bV_{ig}\right\}-\tr\left\{\hat\bLambda_g\bV_{ig}\right\}\right)\right].
\end{align*}
On taking the derivative with respect to $\bLambda_g$, we obtain the update
\begin{equation}  \label{VVE01}
\hat{\bA}_g^{\new}=\frac{\hat{\beta}_g^{\new}}{n_g} \sum_{i=1}^N \tau_{ig}\tr\left\{\left(\hat{\bA}_g\right)^{-1}\bV_{ig}^{\new}\right\}^{\hat{\beta}_g^{\new}-1} \bV_{ig}.
\end{equation}\\

$\hat\beta_g^{\new}\in[1,\infty)$: Using the Jordan decomposition, we write $\bSigma_g^{-1}=\bD\bA_g^{-1}\bD'$. Now, let $\bSigma_g^{-1}=\bD\bLambda_g^{1/\hat{\beta}_g^{\new}}\bD'$, where $\bLambda_g^{-1/\hat{\beta}_g^{\new}}=\bA_g$. Proceeding in a similar fashion to the $\bA_g^{\new}$ update in the VVV ($\hat\beta_g^{\new} \in(1,\infty)$) case, we can get the update
\begin{equation} \label{VVE1inf}
\hat{\bA}_g^{\new}=\left(\frac{\hat{\beta}_g^{\new}}{n_g}\sum_{i=1}^N \tau_{ig}\hat{\bA}_g^{\frac{\hat{\beta}_g^{\new}}{2}} \bW_{ig}^{\hat{\beta}_{g}^{\new}} \hat{\bA}_g^{\frac{\hat{\beta}_g^{\new}}{2}}\right)^{1/\hat{\beta}_g^{\new}},
\end{equation}
where $\bW_{ig}=\hat{\bA}_g^{-\frac{1}{2}} \bV_{ig} \hat{\bA}_g^{-\frac{1}{2}}$, $\bv_{ig}=\hat{\bD}' \bm_{ig}$ and $\bV_{ig} = \bv_{ig} \bv'_{ig}$.\\

The update for $\bD^{\new}$, i.e., $\bD_g$ constrained to be equal across groups (same as $\bGamma$ in Table \ref{tab:models}), is similar to the update for $\bD_g^{\new}$ in the VVV model. We again use an accelerated line search for optimization on the orthogonal Stiefel manifold as employed by \cite{browne2014}. Let $\bQ_{g}=\sum_{i=1}^N \tau_{ig}^{{1}/{\hat\beta_g^{\new}}}\bM_{ig}^{\new}$. The objective function that needs to be minimized is 
$ f(\bD)= \sum_{g=1}^{G} \tr \left\{\bQ_g\bD\left(\hat{\bA}_g^{\new}\right)^{-1}\bD'\right\}^{\hat\beta_g^{\new}},$
with an unconstrained gradient $$ \bar{\text{grad}f(\bD)}= \sum_{g=1}^{G}2\hat\beta_g^{\new}\left(\bQ_g\bD\left(\hat{\bA}_g^{\new}\right)^{-1}\bD'\right)^{(\hat\beta_g^{\new}-1)}\bQ_g\bD\left(\hat{\bA}_g^{\new}\right)^{-1}=\mathbf{R}_g .$$ As shown in \cite{browne2014}, the direction of the steepest descent while in $T_{\bX}\mathcal{M}$ (the tangent space of $\bX$) at the position $\bX$ is
$\text{grad} f(\bX)  = \mathbf{P}_{\bX} \left(\bar{\text{grad}f(\bX)}\right),$ where $\mathbf{P}_{\bX} \left(\bZ\right)=\bZ-\bX \frac{(\bX'\bZ+\bZ'\bX)}{2}$ is the orthogonal projection $\mathbf{P}_{\bX}$ of a matrix $\bZ$ onto $T_{\bX}\mathcal{M}$. Hence, we get 
$$\text{grad} f(\bD) = \sum_{g=1}^{G}\mathbf{R}_g -\frac{1}{2} \sum_{g=1}^{G}\bD\mathbf{R}_g'\bD-\frac{1}{2} \sum_{g=1}^{G}\bD\bD'\mathbf{R}_g.$$
To obtain convergence, the step size $t^*$ is taken to be the Armijo step size (which guarantees convergence) and $\bD$ is updated as
\begin{equation} 
\hat{\bD}^{\new}=\mathbf{R}_{\bX}\left[ -t^*_k \times \text{grad} f(\hat{\bD}) \right], \label{VVE1inforth}
\end{equation} 
where $\mathbf{R}_{\bX}$ is a retraction $\mathbf{R}$ at $\bX$. As before, we use the QR decomposition-based retraction, similar to \cite{browne2014}.

\subsection{Scale structure VII}
Recall that the VII scale structure refers to an isotropic constraint such that $\bSigma_{g}=\lambda_g \mathbf{I}_p$. Then, on ignoring terms not involving $\bSigma_g$, we have
\begin{equation*}
\mathcal{Q} (\lambda_g)=\sum_{i=1}^N\sum_{g=1}^G\frac{\tau_{ig}}{2} \log|{\lambda_g \bI_p}|^{-1}-\frac{\tau_{ig}}{2}  \left[  \left(\bx_i-\hat{\bmu}_g^{\new}\right)'(\lambda_g \bI_p)^{-1} \left(\bx_i-\hat{\bmu}_g^{\new}\right) \right]^{\hat\beta_g^{\new}} .
\end{equation*}
Setting the derivative with respect to $\lambda_g^{-1}$ to $0$ yields $$\lambda_g p n_g - \hat\beta_g^{\new} \sum_{i=1}^N \tau_{ig} \lambda_g^{1-\hat\beta_g^{\new}}\{(\bx_i-\hat{\bmu}_g^{\new})'(\bx_i-\hat{\bmu}_g^{\new})\}^{\hat\beta_g^{\new}}=0.$$ Hence,
\begin{equation} \label{VII}
\hat{\lambda}_g^{\new}=\left(\frac{\hat{\beta}_g^{\new}}{pn_g}\sum_{i=1}^N \tau_{ig}\left[\left(\bx_i-\hat{\bmu}_g^{\new}\right)'\left(\bx_i-\hat{\bmu}_g^{\new}\right)\right]^{\hat{\beta}_g^{\new}}\right)^{1/\hat{\beta}_g^{\new}}.
\end{equation}

\subsection{Scale structure EII}
Recall that the EII scale structure refers to an isotropic constraint such that $\bSigma_{g}=\bSigma=\lambda \mathbf{I}_p$. Then, $$\mathcal{Q} (\lambda)=\sum_{i=1}^N\sum_{g=1}^G\frac{\tau_{ig}}{2} \log|{\lambda \bI_p}|^{-1} - \frac{\tau_{ig}}{2}  \left[ \left(\bx_i-\hat{\bmu}_g^{\new}\right)'(\lambda \bI_p)^{-1}\left(\bx_i-\hat{\bmu}_g^{\new}\right) \right]^{\hat\beta_g^{\new}} .$$
Setting the derivative with respect to $\lambda^{-1}$ to $0$ yields $$pN\hat{\lambda}^{\new} - \sum_{g=1}^G \hat{\beta}_g^{\new} \left(\hat{\lambda}^{\new}\right)^{1-\hat{\beta}_g^{\new}} \sum_{i=1}^N \tau_{ig} \{(\bx_i - \hat{\bmu}_g^{\new})' (\bx_i - \hat{\bmu}_g^{\new})\}^{\hat{\beta}_g^{\new}}=0.$$ Hence, $\hat{\lambda}^{\new}$ can be found by solving the equation
\begin{equation} \label{EII}
pN=\sum_{g=1}^G \hat{\beta}_g^{\new} \left(\hat{\lambda}^{\new}\right)^{-\hat{\beta}_g^{\new}}\sum_{i=1}^N \tau_{ig} \left[\left(\bx_i-\hat{\bmu}_g^{\new}\right)'\left(\bx_i-\hat{\bmu}_g^{\new}\right)\right]^{\hat{\beta}_g^{\new}}. 
\end{equation}

\subsection{Scale structure EEE} \label{sec:EEE}
Here, we provide details on estimation of the scale matrix when it is constrained between groups. Denote $\bM_{ig}^{\new}=\bm_{ig} \bm'_{ig}$, where $\bm_{ig}=\bx_i-\hat{\bmu}^{\new}_g$. On ignoring terms not involving $\bSigma$, 
\begin{equation*}
\mathcal{Q} (\bSigma)=\sum_{i=1}^N\sum_{g=1}^G\frac{\tau_{ig}}{2} \log|{\bSigma}|^{-1}-\frac{\tau_{ig}}{2}  \left[ \bm_{ig}'\bSigma^{-1}\bm_{ig} \right]^{\hat\beta_g^{\new}} .
\end{equation*}
The updates differ based on the current value of $\hat\beta_g$.

$\forall g\in(1 \ldots G)$ $\hat\beta_g^{\new}\in(0,1)$: Using the Jordan decomposition, we can write  $\bSigma^{-1}=\bD\bA^{-1}\bD'=\bD \bLambda \bD'$, where $\bD$ is an orthonormal matrix, $\bA$ is a diagonal matrix of eigenvalues, and $\bLambda=\bA^{-1}$. We obtain updates for both $\hat{\bA}^{\new}$ and $\hat{\bD}^{\new}$. Using similar ideas as before, we can construct a surrogate function:
\begin{align*}
\tr\left\{\bLambda\bV_{ig}\right\}^{\beta_g} \leq \tr\left\{\hat\bLambda\bV_{ig}\right\}^{\beta_g} +\beta_g\tr\left\{\bv'_{ig}\hat\bLambda\bv_{ig}\right\}^{\beta_g-1} \left[ \tr\left\{\bLambda \bV_{ig}\right\}-\tr\left\{\hat\bLambda\bV_{ig} \right\}\right],
\end{align*}
where $\bv_{ig}=\hat\bD' \hat\bm_{ig}$, $\bV_{ig} = \bv_{ig} \bv_{ig}'$. 
Then, using the above, an estimate can easily be obtained
\begin{equation}  \label{EEE01}
\hat{\bA}^{\new}=\frac{1}{N}\sum_{g=1}^G\hat{\beta}_g^{\new} \sum_{i=1}^N \tau_{ig}\tr\left\{\hat{\bA}^{-1}\bV_{ig}\right\}^{\hat{\beta}_g^{\new}-1} \bV_{ig}.
\end{equation}\\

$\exists g\in(1,\ldots,G)$ such that $\hat\beta_g^{\new}\in[1,\infty)$: Let $\bSigma_g^{-1}=\bD\bA^{-1}\bD'=\bD\bLambda^{1/\beta^*}\bD'$, where $\bLambda^{-1/\beta^*}=\bA$, $\beta^*=\text{max}(\beta_1,\ldots,\beta_G)$ and $\beta^*\geq1$. Note that $$\tr\{\bLambda^{1/\beta^*}\bV_{ig}\}^{\beta_g} = \left(\sum_{h=1}^p \lambda_{h}^{1/\beta^*} v_{igh}^2\right)^{\beta_g},$$ where $\bLambda=\mbox{diag}(\lambda_{1},\ldots,\lambda_{p})$. This function is concave with respect to the eigenvalues $\blambda$ (similar to a composition of a weighted $p$-norm and a variable raised to a power less than or equal to 1). Then, the following update can be obtained by proceeding in a similar fashion to the VVV case:
\begin{equation} \label{EEE1infdiag}
\hat{\bA}^{\new}=\frac{1}{N} \sum_{g=1}^G \left(\hat{\beta}_g^{\new}\sum_{i=1}^N \tau_{ig}\hat{\bA}^{\frac{(\hat{\beta}^{*})^{\new}}{2}} \bW_{ig}^{\hat{\beta}_g^{\new}} \hat{\bA}^{\frac{(\hat{\beta}^{*})^{\new}}{2}}\right)^{1/(\hat{\beta}^{*})^{\new}},
\end{equation}
where $\bv_{ig}=\hat{\bD}' \bm_{ig}$, $\bV_{ig}=\bv_{ig}\bv_{ig}'$, 
and $\bW_{ig}=\hat{\bA}^{-\frac{1}{2}} \bV_{ig} \hat{\bA}^{-\frac{1}{2}}$.\\

The update for $\bD$ is similar to the update for the VVE model. Let $$\bQ_{g}=\sum_{i=1}^N \tau_{ig}^{{1}/{\hat\beta_g^{\new}}}\bM_{ig}^{\new}$$ and the objective function that needs to be minimized now is 
$$ f(\bD)= \sum_{g=1}^{G} \tr \left\{\bQ_g\bD\left(\hat\bA^{\new}\right)^{-1}\bD'\right\}^{\hat\beta_g^{\new}}.$$

\subsection{Scale structure EEI} \label{sec:EEI}

The estimate for the diagonal matrix of eigenvalues $\bSigma$ in the EEI case can be derived using ideas similar to the EEE and VVI case. We obtain

$\forall g\in(1 \ldots G)$ $\hat\beta_g^{\new}\in(0,1)$:
\begin{equation}  \label{EEI01}
\hat{\bSigma}^{\new}=\frac{1}{N}\sum_{g=1}^G\hat{\beta}_g^{\new} \sum_{i=1}^N \tau_{ig}\tr\left\{\hat{\bSigma}^{-1}\bM_{ig}^{\new}\right\}^{\hat{\beta}_g^{\new}-1} \bM_{ig}^{\new}.
\end{equation}
\\

$\exists g\in(1,\ldots,G)$ such that $\hat\beta_g^{\new}\in[1,\infty)$:

\begin{equation} \label{EEI1infdiag}
\hat{\bSigma}^{\new}=\left(\frac{1}{N} \sum_{g=1}^G \hat{\beta}_g^{\new}\sum_{i=1}^N \tau_{ig}\hat{\bSigma}^{\frac{(\hat{\beta}^{*})^{\new}}{2}} \bW_{ig}^{\hat{\beta}_{g}^{\new}} \hat{\bSigma}^{\frac{(\hat{\beta}^{*})^{\new}}{2}}\right)^{1/(\hat{\beta}^{*})^{\new}},
\end{equation}
where $\bW_{ig}=\hat{\bSigma}^{-\frac{1}{2}} \bM_{ig}^{\new} \hat{\bSigma}^{-\frac{1}{2}}$. 
\\

\subsection{Scale structure EEV} \label{sec:EEV}

The estimate for $\bSigma_g$ in the EEV case can be derived using ideas similar to the EEE and VVV case. 

$\forall g\in(1 \ldots G)$ $\hat\beta_g^{\new}\in(0,1)$: 
\begin{equation}  \label{EEV01}
\hat{\bA}^{\new}=\frac{1}{N}\sum_{g=1}^G\hat{\beta}_g^{\new} \sum_{i=1}^N \tau_{ig}\tr\left\{\hat\bA^{-1}\bV_{ig}\right\}^{\hat{\beta}_g^{\new}-1} \bV_{ig},
\end{equation}
where $\bv_{ig}=\hat\bD_g' \bm_{ig}$, $\bV_{ig}=\bv_{ig}\bv_{ig}'$. 
\\

$\exists g\in(1,\ldots,G)$ such that $\hat\beta_g^{\new}\in[1,\infty)$: 

\begin{equation} \label{EEV1infdiag}
\hat{\bA}^{\new}=\frac{1}{N} \sum_{g=1}^G \left(\hat{\beta}_g^{\new}\sum_{i=1}^N \tau_{ig}\hat\bA^{\frac{(\hat{\beta}^{*})^{\new}}{2}} \left(\bW_{ig}\right)^{\hat{\beta}_{g}^{\new}} \hat{\bA}^{\frac{(\hat{\beta}^{*})^{\new}}{2}}\right)^{1/(\hat{\beta}^{*})^{\new}},
\end{equation}
where $\bv_{ig}=\hat\bD_g' \bm_{ig}$, $\bV_{ig}=\bv_{ig}\bv_{ig}'$, 
and $\bW_{ig}=\hat\bA^{-\frac{1}{2}} \bV_{ig} \hat\bA^{-\frac{1}{2}}$.\\

The update for $\bD$ is similar to the EEE and VVV models. Let $\bQ_{g}=\sum_{i=1}^N \tau_{ig}^{{1}/{\hat\beta_g^{\new}}}\bM_{ig}^{\new}$. The objective function that needs to be minimized now is 
$$f(\bD_g)= \sum_{g=1}^{G} \tr\left\{\bQ_g\bD_g\left(\hat\bA^{\new}\right)^{-1}\bD'_g\right\}^{\hat\beta_g^{\new}}.$$

\section{Initialization, model selection, and performance assessment}\label{sec:modelspecifications}
\subsection{Model selection and initialization}\label{sec:initialization}
In model-based clustering applications, it is common to fit each member of a family of mixture models for a range of values of $G$, out of which a `best' model is chosen based on some likelihood-based criterion. Note that this best model does not necessarily correspond to optimal clustering. The Bayesian information criterion \citep[BIC;][]{schwarz1978} is commonly used for mixture model selection. Even though the regularity properties needed for the development of the BIC are not satisfied by mixture models \citep{keribin1998, keribin2000}, it has been used extensively \citep[e.g.,][]{dasgupta1998, fraley2002} and performs well in practice. The BIC can be computed as 
$$\text{BIC}=2l(\hat{\bTheta})-m\log{N},$$ 
where $l(\hat{\bTheta})$ is the maximized log-likelihood, $m$ is the number of free parameters, and $N$ is the sample size. The integrated completed likelihood \citep[ICL;][]{biernacki2000} aims to correct the BIC by putting some focus on the clustering performance. This is done via the estimated mean entropy, which reflects the uncertainty in the classification of observations into components. The ICL can be computed via
$$\text{ICL}\approx\mbox{BIC}+\sum_{i=1}^{N}\sum_{g=1}^{G}\mbox{MAP}(\tau_{ig})\log{\tau_{ig}},$$ where $\mbox{MAP}(\tau_{ig})$ is the maximum \emph{a posteriori} probability, equaling 1 if $\mbox{max}_{h=1,\ldots,G}(\tau_{ih})$ occurs at component $h=g$, and 0 otherwise.

Because the EM algorithm is iterative, initial values are needed for the parameters. The issue of starting values is important because the performance of the EM algorithm is known to depend on the starting values. Poor starting values can result in singularities or convergence to local maxima \citep{titterington1985}. Some techniques that can alleviate such issues are constraining eigenvalues \citep{ingrassia2007, browne2013constrained}, deterministic annealing \citep{zhou2010}, or picking a run from multiple starts for the EM. The algorithm can be initialized based on a random assignment of data points to components, on $k$-means clustering \citep{hartigan1979}, on some hierarchical clustering method, or in some other way. 
We constrain $\beta_g$ to be less than 200 for numerical stability---this is similar to how the degrees of freedom parameter in mixtures of $t$-distributions is sometimes constrained to be less than 200 \citep{andrews2011b}.

\subsection{Convergence criterion}
Here, a stopping criterion based on Aitken's acceleration \citep{aitken1926} is used to determine convergence. The commonly used lack of progress criterion can converge earlier than the Aitken's stopping criterion, resulting in estimates that might not be close to the maximum likelihood estimates.  The Aitken acceleration at iteration $k$ is $$a^{(k)}=\frac{l^{\new}-l^{(k)}}{l^{(k)}-l^{(k-1)}},$$ 
where $l^{(k)}$ is the log-likelihood value from iteration $k$. An asymptotic estimate of the log-likelihood at iteration $k+1$ can be computed via $$l_{A}^{\new}=l^{(k)}+\frac{1}{1-a^{(k)}}(l^{\new}-l^{(k)})$$ \citep{bohning1994}. Convergence is assumed to have been reached when $l_A^{\new}-l^{k}<\epsilon$, provided that this difference is positive \citep[cf.][]{lindsay1995,mcnicholas2010a}. Note that we use $\epsilon=0.005$ herein.
 
\subsection{Performance assessment}
The adjusted Rand index \citep[ARI;][]{hubert1985} is used for determining the performance of the chosen model by comparing predicted classifications to true group labels, when known. The ARI corrects the Rand index \citep{rand1971} to account for chance when calculating the agreement between true labels and estimated classifications. An ARI of 1 corresponds to perfect agreement, and the expected value of the ARI is 0 under random classification. \cite{steinley2004} provides a thorough evaluation of the ARI.


%
%
%

\begin{table*}[!ht]
\caption{Time taken in seconds to run all sixteen models (based on un-optimized code) for the real data examples for $G=1,\ldots,5$.} \label{timing}
\begin{tabular*}{1.0\textwidth}{@{\extracolsep{\fill}}lr}
\hline
Data & Time taken (seconds)\\
\hline
{\tt body} ($p=24$, $G=2$, $N=507$) & 19151\\
{\tt diabetes} ($p=3$, $G=3$, $N=145$) & 310\\
{\tt female voles} ($p=7$, $G=2$, $N=86$) &  291\\
{\tt wine} ($p=13$, $G=3$, $N=178$) &  2326\\
{\tt srbct} ($p=10$, $G=4$, $N=83$) &  1101\\
{\tt golub} ($p=10$, $G=2$, $N=72$) & 405\\
\hline
\end{tabular*}

\bigskip
Dimensionality, the number of known groups (i.e., classes), and the number of sample points are in parenthesis following the name of each data set.
\end{table*}

\end{document}